\documentclass[a4paper,fleqn,usenatbib]{mnras}

\usepackage[T1]{fontenc}
\usepackage{ae,aecompl}
\usepackage{natbib}
\usepackage{lscape}
\usepackage{lastpage}
\usepackage{ulem}
\usepackage{cancel}

\usepackage{graphicx}	
\usepackage{amsmath}	
\usepackage{amssymb}	
  \usepackage{multicol}        
   \usepackage{bm}		
   \usepackage{pdflscape}	
    \usepackage{subfig}
 \usepackage{enumerate}
 \usepackage{rotating}
\usepackage{float}
\usepackage{adjustbox}
\usepackage{hyperref}
\usepackage{color,soul}
\usepackage[flushleft]{threeparttable}
\usepackage{comment}
\usepackage{soul}

\newcommand{\kms}{\,km\,s$^{-1}$} 

\newcommand{\Msol}{M$_{\odot}$}                           
\newcommand{\Lsol}{L$_{\odot}$} 				

\newcommand{\hii}{H\,{\sc ii }\rm}

\newcommand{\oiii}{[O\,{\sc iii}]}
\newcommand{\oii}{[O\,{\sc ii}]}
\newcommand{\oi}{[O\,{\sc i}]}
\newcommand{\nii}{[N\,{\sc ii}]}
\newcommand{\sii}{[S\,{\sc ii}]}
\newcommand{\siii}{[S\,{\sc iii}]}
\newcommand{\neiii}{[Ne\,{\sc iii}]}
\newcommand{\hei}{He\,{\sc i}}
\newcommand{\hi}{H\,{\sc i}}

\newcommand{\heii}{He\,{\sc ii}}
\newcommand{\feiii}{[Fe\,{\sc iii}]}
\newcommand{\ariv}{[Ar\,{\sc iv}]}

\newcommand{\ho}{$\itl H_{0}$}

\newcommand{\kmsmpc} {$\rm {km~s^{-1}}~Mpc^{-1}$}

\newcommand{\usfr}{M$_{\odot}$yr$^{-1}$}
\def\ulum{erg $\rm s^{-1}$}
\def\uflux{erg $\rm cm ^{-2} \rm s^{-1}$}

\def\hepsi {H$\epsilon$}
\def\hdelta {H$\delta$}
\def\hgamma {H$\gamma$}
\def\hbeta {H$\beta$}
\def\halpha {H$\alpha$}
\def\eqwhbe{ EW(H$\beta$)}
\def\eqwhal{ EW(H$\alpha$)}

\def\gb{{\it g}}

\def\oh{12+$\log$(O/H)}
\def\nh{12+$\log$(N/H)}
\def\he{12+$\log$(He$^+$/H$^+$)}
\def\ohmas{12+$\log$(O$^{+}$/H$^{+}$)}
\def\ohmasmas{12+$\log$(O$^{++}$/H$^{+}$)}

\def\lum{L(H$\beta$)}

\newcounter{mycount}

\author[D. Fern\'andez Arenas ] {D. Fern\'andez-Arenas$^{1}$\thanks{E-mail: arenas@inaoep.mx}, E. Carrasco$^{1}$, R. Terlevich$^{1,2}$, E. Terlevich$^{1}$\thanks{Visitor at IoA and KICC, Cambridge, UK}, R. Amor\'in$^{3,4}$,
\newauthor F. Bresolin$^{5}$,  
R. Ch\'avez$^{6,7}$,  
A. L. Gonz\'alez-Mor\'an$^{8,9}$, 
D. Rosa-Gonz\'alez$^{1}$, 
Y. D. Mayya$^{1}$,
\newauthor O. Vega$^{1}$,
J. Zaragoza-Cardiel$^{1,7}$,
J. M\'{e}ndez-Abreu$^{8,9}$,
R. Izazaga-Pérez$^{1}$,
A. Gil de Paz$^{10,11}$,
\newauthor  J. Gallego$^{10,11}$,
J. Iglesias-P\'{a}ramo$^{12}$,
M.L. Garc\'{i}a-Vargas$^{13}$,   
 P. G\'{o}mez-Alvarez$^{13}$,  
\newauthor A. Castillo-Morales$^{10, 11}$,
N. Cardiel$^{10,11}$,
S. Pascual$^{10,11}$,
A. P\'{e}rez-Calpena$^{13}$.\\
$^{1}$ Instituto Nacional de Astrof\'\i sica, \'Optica y Electr\'onica,Tonantzintla, Puebla, Mexico \\
$^{2}$ Institute of Astronomy, University of Cambridge, Cambridge, CB3 0HA, UK\\
$^{3}$ Departamento de Astronom\'ia, Universidad de La Serena, Av. Juan Cisternas 1200 Norte, La Serena, Chile\\
$^{4}$ Instituto de Investigaci\'on Multidisciplinar en Ciencia y Tecnolog\'ia, Universidad de La Serena, Ra\'ul Bitr\'an 1305, La Serena, Chile\\
$^{5}$ Institute for Astronomy, 2680 Woodlawn Drive, Honolulu, HI 96822, USA\\
$^{6}$ Instituto de Radioastronom\'ia y Astrof\'isica, UNAM, Campus Morelia, C.P. 58089, Morelia, M\'exico. \\
$^{7}$ Consejo Nacional de Ciencia y Tecnolog\'ia, Av. Insurgentes Sur 1582, 03940 Mexico City, Mexico\\
$^{8}$ Instituto de Astrof\'isica de Canarias, C/ V\'ia L\'actea s/n, 38205 La Laguna, Tenerife, Spain\\
$^{9}$ Departamento de Astrofísica, Universidad de La Laguna, E-38205 La Laguna, Tenerife, Spain.\\
$^{10}$ Departamento de F{\'\i}sica de la Tierra y Astrof{\'\i}sica, Fac. CC. F{\'\i}sicas,\\
Universidad Complutense de Madrid, Plaza de las Ciencias, 1, E-28040 Madrid, Spain \\
$^{11}$ Instituto de F{\'\i}sica de Part{\'\i}culas y del Cosmos, IPARCOS, Fac. CC. F{\'\i}sicas, Universidad Complutense de Madrid, \\
Plaza de las Ciencias 1, E-28040 Madrid, Spain\\
$^{12}$ Instituto de Astrof{\'\i}sica de Andaluc{\'\i}a, IAA-CSIC, Glorieta de la Astronom{\'\i}a s/n, E-18008  Granada, Spain\\
$^{13}$ FRACTAL S.L.N.E. Calle Tulip{\'a}n 2, portal 13, 1A, E-28231 Las Rozas de Madrid, Spain \\
}

\title[Spatially-resolved properties of the ionized gas in the HII galaxy...]{Spatially-resolved properties of the ionized gas in the HII galaxy J084220+115000}

\begin{document}

\setcounter{mycount}{0}

\label{firstpage}
\pagerange{\pageref{firstpage}--\pageref{lastpage}}
\maketitle

\begin{abstract}

We present a spatially resolved spectroscopic study for the metal poor \hii galaxy J084220+115000 using MEGARA Integral Field Unit observations at the Gran Telescopio Canarias. We estimated the gas metallicity using the direct method for oxygen, nitrogen and helium and found a mean value of \oh=$8.03\pm$0.06, and integrated electron density and temperature of $\sim161$ cm$^{-3}$ and  $\sim15400$ K, respectively. The metallicity distribution shows a large range of $\Delta$(O/H) = 0.72 dex between the minimum and maximum (7.69$\pm$0.06 and 8.42$\pm$0.05) values, unusual in a dwarf star-forming galaxy. We derived an integrated $\log$(N/O) ratio of $-1.51\pm0.05$ and found that both  N/O and  O/H correspond to a primary production of metals. Spatially resolved maps indicate that the gas appears to be photoionized by massive stars according to the diagnostic line ratios. Between the possible mechanisms to explain the starburst activity and the large variation of oxygen abundance in this galaxy, our data support a possible scenario where we are witnessing an ongoing interaction triggering multiple star-forming regions localized in two dominant clumps.

\end{abstract}
 

\section{Introduction}

Observationally HII Galaxies (HIIG)\footnote{``Peculiar" objects from Zwicky list of compact galaxies were identified by \cite{Sargent1970a,Sargent1970b} and further studied by \cite{Sargent1970}(SS1970). Those that have spectral characteriscs indistinguishable from giant extragalactic HII regions (like 30~Dor in the LMC) were designated by SS1970 as ``Isolated Extragalactic HII Regions". Later on, these compact, dwarf, star forming galaxies, were dubbed HII Galaxies \citep{cam86,ter91,Maza1991}}. and Giant HII Regions (GHIIRs) represent the youngest Super Star Clusters (SSCs)  that can be observed in any detail. In particular, the study of HIIGs provides important clues about the intrinsic properties of young or unevolved galaxies. These clues are important in searches for primaeval galaxies, particularly if starbursts are the dominant mode of galaxy formation. Though their optical-UV light is clearly dominated by young stars, HIIGs are unlikely to be truly ``young" in the sense of completely lacking old stars \cite[e.g][]{Telles1997a}. This was found even in the most extreme low abundance objects like IZw18 \citep[e.g][]{Aloisi2007,Annibali2013}.

HIIGs have a high specific star formation rate  \citep[\mbox{$ \log({\rm sSFR)}\sim -7 \;{\rm to}-6.5\; {\rm yr}^{-1}$},][]{searle1972,Telles2018}. Studies indicate that the recent star formation is concentrated in extreme SSCs with sizes of $\sim$ 20 pc \citep[e.g.][]{telles2003} and low heavy-element abundance  (1/50 < Z < 1/3 Z$_{\odot}$). In fact, the most metal-poor compact starbursts at all redshifts tend to appear as HIIGs \citep{Kunth2000,GildePaz2003,Amorin2012,Izotov2012,Kehrig2016,Amorin2017,Kehrig2018,Wofford2021}.

Most HIIGs were discovered in objective-prism surveys thanks to their strong narrow emission lines. Currently, in spectroscopic surveys like the Sloan Digital Sky Survey (SDSS), they are selected by very large equivalent widths in the Balmer lines. Since the luminosity of HIIGs is dominated by the starburst component they can be observed even at large redshifts becoming interesting standard candles. \cite[cf.][]
{melnick2000,plionis2011,terlevich2015,chavez2016,Yennapureddy2017,Ruan2019,gonzalez2019,wu2020,Gonzalez2021,Tsiapi2021}.

Many of the properties of HIIGs are usually obtained from single aperture or long slit observations. The data are then used to derive the physical conditions of the gas (temperatures and densities) and to estimate abundances and ionization conditions, as well as characteristics of the ionizing star clusters (e.g. mass, age, effective temperature). These results constitute the main body of our knowledge of the conditions of the gas in HIIGs and GHIIRs \citep[e.g.][]{perezmontero2005,hagele2007,perezmontero2014}. 

The assumption is that the measurements derived from single aperture or long-slit are representative of the whole nebular conditions and variations within the nebula are assumed to be minimal or non existent. This scenario can also be supported by the fact that in general, these type of studies do not find a significant abundance gradient or large deviation in kinematics properties in the ionized gas of HIIGs or GHIIRs even for HIIGs showing multiple star-forming knots \citep{Esteban1995,Vilchez1996,Vilchez2003,papaderos2006,Perez2009,perezmontero2011,Esteban2020}. 

Integrated observations, such as long-slit or single fiber, may fail to correlate the spatial location of star-forming regions with the physical conditions and the chemical abundances of the interstellar medium (ISM) as derived from the optical emission-lines. Traditionally, the physical conditions (age and metallicity of the ionising clusters and the kinematics and metallicity of the ionised gas) of metal-poor HIIGs, have been derived from long slit observations, either centred on the brightest regions or by several observations scanning the whole galaxy. Presently, integral field spectroscopy constitutes a powerful tool to obtain simultaneously information, not only on the brightest condensations but also on the diffuse matter surrounding them. This helps to understand better the interplay between the massive stars population and the properties of the Interstellar Medium in metal-poor galaxies as has been amply illustrated in the literature in recent years
\citep[e.g][]{Lagos2009,Perez2013,Kehrig2016,Kehrig2018,Bosch2019,Kumari2019,James2020,Bresolin2020,Wofford2021,Carrasco2022}.

Our main motivation in this paper is to present a study of the spatially-resolved properties of the ionized gas of the metal-poor HIIG: SDSS J084220.94+115000.2  (hereafter, J0842+1150) that shows multiple bursts of star formation that can be identified in the Hubble Space Telescope (HST) image and  in the Chandra data. This study is based in observations obtained at the  Gran Telescopio Canarias (GTC),  using the integral field unit (IFU) of  the instrument MEGARA (acronym of Multi-Espectrógrafo en GTC de Alta Resolución para Astronomía). GTC is located at the Observatorio del Roque de los Muchachos in la Palma, Canary Island, Spain. 

The paper is organized as follows: In \S \ref{galdescrip}, we present a complete description of the HII galaxy J0842+1150. In  \S \ref{data}, we describe the observations and data reduction. In \S \ref{fluxes}, we report the flux measurements and optical emission line intensity maps. In \S \ref{properties}, we derived from the observations the maps of the properties of the ionized gas. Abundance maps for helium, oxygen and nitrogen derived from the direct method based on determinations of the electron temperature are presented in \S \ref{abundances}. In \S \ref{abundances3}, we compare strong-line methods to derive abundances for the individual spaxels and discuss the nitrogen to oxygen abundance ratio for J0875+1150. In \S \ref{Oinhomogeneities}, we discuss the chemical inhomogeneities in HIIGs and Blue Compact Dwarf galaxy (BCD) for the case of J0842+1150.  Finally, in \S \ref{summaryandconclu}, we present a summary of this work and our main conclusions.

\section{The HII galaxy J0842+1150}\label{galdescrip}

J0842+1150 also known as Cam 0840 + 1201 was reported for the first time by \cite{cam86}, as a low-metallicity HIIG with values of O/H$=0.785\times10^{-4}$ (\oh = 7.89) and N/H$=0.270\times10^{-5}$ (\nh = 6.43) with an electron temperature of $T_e$\oiii=13891 K. Using  International Ultraviolet Explorer (IUE)  observations, \cite{Terlevich1993} detected  Ly$\alpha$ emission in this galaxy.

In the morphological classification scheme for HIIGs  by \cite{telles1997}, J0842+1150  was reported as an HIIG with perturbed and extended morphology, with two dominant giant HII regions and signs of tails beyond the star-forming regions. They reported an oxygen abundance of \oh = 7.88 and a velocity dispersion of 36.5 \kms. J0842+1150 appears to be an interacting galaxy consisting of two clumps of multiple star-forming regions, each one with unresolved X-ray emission,  detected at 0.5-8 keV by \cite{Brorby2017}.

\cite{Kehrig2004}, analysing the chemical abundances in HIIGs with a detailed spectroscopic study, report  abundances of \oh=7.82$\pm$0.08  and $12+\log$(S/H)=6.36$\pm$0.33 with a  $T_e$\oiii\ of 13600$\pm$600 K. \cite{Brinchmann2008} classified it as a Wolf-Rayet (WR) galaxy and also derived an abundance of \oh=8.09 using the direct $T_e$ method and the SDSS spectrum.  \cite{Bordalo2011}  found a metallicity (\oh) of 7.98 and also reported a velocity dispersion in \kms\ of 36.8$\pm$1.8, 34.9$\pm$0.4, 34.6$\pm$0.7 and 34.4$\pm$0.2 for the \hbeta, \halpha\ and \oiii$\lambda\lambda$4959,5007 lines, respectively. Whereas \cite{Chavez2012} reported 32$\pm$3\kms\ for \hbeta\ and 27$\pm$1\kms\ in \oiii. The different values reported in the literature may be due to a pointing problem.

\subsection{Is J0084220 a Lyman Break Analog?}

Interestingly, J0084220 shows similar shapes to those that have been reported in local nearby compact UV-luminous galaxies (UVLGs) that closely resemble high-redshift Lyman Break Galaxies (LBGs), sometimes called ``Lyman Break Analogs'' (LBAs). Analysis of their SDSS spectra and of their spectral energy distributions has shown that the LBAs are similar to LBGs in their basic global properties, thus enabling a detailed investigation of many processes that are important in star-forming galaxies at high redshift. Common characteristics include faint tidal features and UV/optical light  dominated by unresolved ($\sim$100-300 pc) super starburst regions (SSBs), suggesting that the starbursts are the result of a merger or interaction \citep{Grimes2007,Overzier2008,Overzier2009,Goncalves2010,Basu-Zych2009}.

Additional  properties for LBA are stellar masses  as low as 10$^{8.5}$\Msol\ and \oh<8.5 with particular detection of X-ray emission \citep{Brorby2016} which  match the properties of the most luminous HIIGs. Using  the  FUV flux from  \citet{rosa2002}, we derived a $L_{\rm FUV}=10^{10.13}$\Lsol\ and adopting the radius in the $u$-band from the SDSS reported in \cite{Chavez2014}, we derived a $I_{\rm FUV}$ of $10^{9.07}$\Lsol/kpc$^2$. Combined with the morphology, this places J084220+11500 as a potential candidate for LBA which motivates the study of this galaxy and also a larger sample of HIIGs in the context of LBAs and Lyman-$\alpha$ Emitters and the precise connection between them. Thus, J084220+11500 provides a local laboratory to study the extreme star formation processes that could be occurring in high-redshift galaxies.

\begin{figure}
\centering
\includegraphics[scale=0.35]{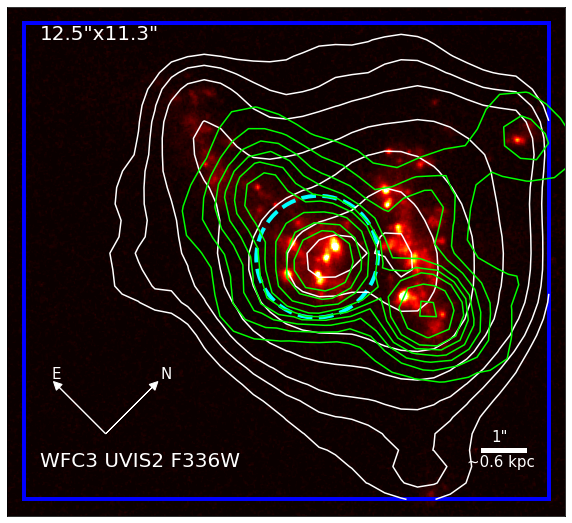}
\includegraphics[scale=0.35]{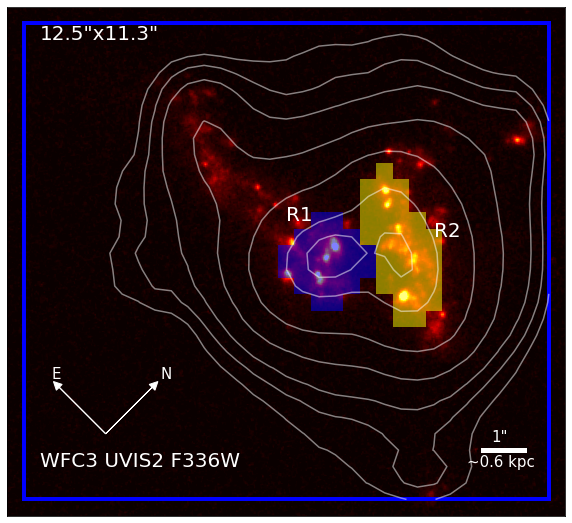}

    \caption[]{HST WFPC3/F336W archival image of J0842+1150 (HST, Proposal ID 13940; PI: Matt Brorby). The blue box denotes MEGARA IFU field of view (12.5$\times$11.3 arcsec$^{2}$). In the top panel, the cyan dashed circle shows the position of the SDSS 3 arcsec fiber diameter. The 0.5-8 keV X-ray contours (from \cite{Brorby2017}) are shown in green.  Bottom panel:  masked regions where spectra were extracted to analyse individual regions (R1,R2). Here and in all subsequent figures, white contours correspond to the \halpha\ emission map obtained in this work.}
\label{fov}
\end{figure}

In Table~\ref{generalproperties}, we  list  the main  properties of J0842+1150. It has to be indicated that the spectroscopic parameters were derived from single aperture data that do not cover the total nebular emission. In Fig. \ref{fov}, we present the HST image of J0842+1150 in the F336W filter, the field of view (FoV) of MEGARA IFU in blue, the SDSS fiber as a cyan dashed circle, the X-ray emission contours (0.5-8 keV) from \cite{Brorby2017}  in green and in white the contours of the flux of the \halpha\ map derived in this work as described in section \ref{fluxes}. Throughout this work, we also present our MEGARA results obtained for integrated regions R1 and R2 corresponding to main star-forming bursts identified by means of the \halpha\ contours; we also present integrated regions according to the metallicity variation (see the details in Appendix \ref{metal_regions}).

\begin{table}
\centering
\caption{J0842+1150 general properties}
\begin{tabular}{|l|c|}
\hline \hline 
Parameter  & Value  \\ 
\hline 
R.A (J2000)  & 08$^{\rm h}$42$^{\rm m}20$\fs90\\
DEC. (J2000) & +11\degr50\arcmin00\farcs2\\
Redshift    & 0.029$^{a}$\\
D$_L$(Mpc)  & 125.2$^{b}$ \\
Metallicity (\oh) & 8.09$\pm$0.01$^c$\\
SFR (\usfr)   &     7.2$^d$\\
$L_{\rm X} (10^{39}$ \ulum) &36$\pm$ 7.5$^e$\\
SDSS $u$ &17.07$^f$\\
SDSS $r$ &16.63$^f$\\
$\log$L(\hbeta)(\ulum) & 40.85$\pm$0.13$^f$\\
R$_{50}$ ($r$ band)  & 2.81\arcsec $\sim1.65$ kpc $^f$ \\
\hline 
\multicolumn{2}{@{}p{8cm}@{}}{$^a$ Spectroscopic redshift as reported by SDSS. $^b$  Luminosity distance in Mpc (from SDSS redshift). The Hubble Constant adopted throughout this work is \ho=71.0 \kmsmpc, assuming a flat Universe and a $\Lambda$CDM cosmological model with $\Omega_m=0.3$. $^c$ Oxygen abundance derived using the $T_e$-method  and the Sloan fiber \citep{Brinchmann2008}. $^d$ Star formation rate (SFR) from UV+IR. $^e$ Total X-ray luminosity (L$_X$) refers to the 0.5-8 keV band flux assuming a photon index of $\Gamma = 1.7$ reported by \cite{Brorby2017}. $^f$ Parameters as reported or derived from SDSS.}\\
\end{tabular} 
\label{generalproperties}
\end{table}

\section{Observations and data reduction}\label{data}

\subsection{Observations}
Spectroscopic observations of J0842+1150 were carried out using the IFU of MEGARA at the 10.4m  GTC. MEGARA provides integral-field and multi-object spectrograph at low, medium and high spectral resolution $R_{\rm FWHM}$\footnote{$R_{\rm FWHM}\equiv\lambda/\Delta\lambda$ with $\Delta\lambda=$ individual lines Full Width at Half Maximum (FWHM).}$\sim6000,12000$ and $20000$, respectively for the LR, MR and HR modes in the visible wavelength interval covering from 3650 to 9700 \AA, through 18 spectral configurations: six in LR, 10 in MR and two in HR.

In particular, we used the integral field unit mode. MEGARA IFU data consists of 623 spectra, of which 567 are on-object which provide a FoV of 12.5$\times$11.3 arcsec$^2$. Fifty-six fibres are reserved for the sky (8 mini-bundles of 7 fibers each)  located at the edge of the field at distances from 1.7 arcmin to 2.5 arcmin from the centre of the IFU designed to perform simultaneous sky subtraction. The observations were carried out with a rotated FoV of 45$^{\circ}$ in order to minimize the contamination of background galaxies on the sky bundles.

The data were obtained using three different gratings in order to cover the main optical emission lines. LR-U, LR-B and LR-R; the covered wavelength and resolving power of each grating are presented in Table \ref{obs}. The FoV of MEGARA corresponds to 7.2 kpc$\times$6.5 kpc at the distance of the galaxy (125.2 Mpc, see Table~\ref{generalproperties} and Fig. \ref{fov}), with spatial sampling of 0.62 arcsec per spaxel \footnote{This size corresponds to the diameter of the circle on which the hexagonal spaxel is inscribed ($\sim 355$pc).}.

A total of 1.75 hours on the galaxy in dark sky and spectrophotometric conditions were observed, with the integration times split in three exposures for each gratings as indicated in Table \ref{obs}. The seeing was about 1 arcsec during the observations in each spectral configuration, corresponding to different volume phase holographic gratings (VPH) and all science frames were obtained at  similar airmasses close to unity. Additionally, all necessary calibration frames were acquired i.e. spectrophotometric standard stars, halogen lamp flats, ThNe arcs and a series of bias frames.

\subsection{Data reduction}

The data reduction procedure was carried out using \texttt{MEGARA Data Reduction Pipeline (DRP)}\footnote{\url{https://github.com/guaix-ucm/megaradrp}} publicly available and open source under GPLv3+ (GNU Public License, version 3 or later). The DRP is a custom-made user-friendly tool formed by a set of processing recipes developed in \texttt{Python}  \citep{Pascual2019} which is based on a series of processing recipes and the cookbook\footnote{\url{http://doi.org/10.5281/zenodo.3834345}}. The recipes used for obtaining the calibration images were \texttt{MegaraBiasImage, MegaraTraceMap, MegaraModelMap, MegaraArcCalibration, MegaraFiberFlatImage,  MegaraLcbStdStar.} 

Briefly, the data reduction starts generating a Master Bias with the \texttt{MegaraBiasImage} routine from the bias images. Images are corrected from overscan and trimmed to the physical size of the detector. To trace the locus of each of the 623 spectra, the \texttt{MegaraTraceMap} recipe uses the halogen lamp images to find the position of the illuminated fibers on the detector. The \texttt{MegaraModelMap} recipe  takes the results of the previous step and the halogen images as input to produce an optimised extraction of the fiber spectra.  The routine fits simultaneously 623 Gaussians every 200 columns and then interpolates the parameters of the Gaussian for each spectral pixel. With this information, the routine generates a weight map for every fibre which is applied to the data in order to perform the extraction. 

The \texttt{MegaraArcCalibration} recipe uses the lamp wavelength calibration images, their offset value and the output of the \texttt{MegaraModelMap} to produce a wavelength calibration. The \texttt{MegaraFiberFlatImage} recipe is used to correct for the global variations in transmission between fibres and as a function of wavelength. The \texttt{MegaraLcbAcquisition} recipe returns the position of the standard star on the IFU. Once the position of the standard star on the LCB is known, the \texttt{MegaraLcbStdStar} routine produces the Master Sensitivity curve by comparing the 1D flux spectrum of the standard star (corrected from atmospheric extinction) with its tabulated flux-calibrated template. This sensitivity curve also corrects from the spectral instrument response (mostly dominated by VPH transmission and detector quantum efficiency), so that this step is needed even when non-photometric conditions prevent a reliable flux calibration. 

Once all the calibration files are obtained, the science frames  are processed with the recipe \texttt{MegaraLcbImage} producing the Row-Stacked-Spectra (RSS) file with the individual  spectra for all fibres,  wavelength and flux calibrated and corrected for telluric effects.

Finally, we created datacubes with a spaxel scale of 0.4 arcsec in each of the spatial axes, from the RSS file using \texttt{create\_cube\_from\_rss} \footnote{ \texttt{create\_cube\_from\_rss}, available in the  megaradr repository, is a tool written in Python to convert MEGARA reduced dataproducts from the RSS format obtained with megaradrp to a more user-friendly 3D datacube.}.

\begin{table*}
\caption{Observation setup,  November 23rd  2020.}
\begin{tabular}{|l|c|c|c|c|c|}
\hline \hline 

Grating  & Wavelength range & Central wavelength & R$_{\textrm{FWHM}}$  & Exp. time & Airmass  \\ 
             & (\AA)  & (\AA)  &   & (seconds) &   \\ 
\hline 
VPH405-LR (LR-U)	& 3654-4391 & 4025 & 5750 & 900$\times$3 & 1.05 \\
VPH480-LR (LR-B)	& 4332-5199 & 4785 & 5000 & 600$\times$3 & 1.08 \\
VPH675-LR (LR-R)	& 6096-7303 & 6729 & 5900 & 600$\times$3 & 1.09 \\
\hline 
\end{tabular} 
\label{obs}
\end{table*}

\section{Flux measurements and emission line intensity maps}\label{fluxes}

In what follows, we describe the analysis performed to obtain the emission line fluxes. Each observing setup  is characterized by different coverage, central 
wavelength and spectral resolution as described in Table \ref{obs}. We used LR-U  to measure the doublet \oii$\lambda\lambda$3727,29 \AA\ and \hdelta, LR-B for  \hgamma, \oiii$\lambda$4363, \hbeta, \oiii$\lambda\lambda$4959, 5007 and LR-R for  \oi$\lambda$6302, \halpha, \nii$\lambda$6583 and  \sii$\lambda\lambda$6716,6730 \AA.

Emission line fluxes were measured from individual spaxels by fitting single Gaussian curves to the profiles using a least-squares minimization procedure implemented in the \texttt{Python} package \texttt{lmfit} \citep{Newville2014}. For each spaxel, we first averaged the spectrum by considering a box/squared region with a size comparable to the Point Spread Function (PSF) of the observation (2 spaxels). This improves the S/N of the emission line. We fitted a straight line to remove a local continuum using two spectral windows at both sides of the emission lines; those emission line free regions  were used also to measure the RMS. 

To consider the uncertainties associated with the emission line fluxes, we generated a synthetic emission line adding randomly the noise of the continuum (RMS) to the observed spectra and repeated the fit; for each spaxel  200 iterations were performed to estimate the final errors. Finally, we exclude spaxels that have lines with peak signal-to-noise (S/N, with  S the amplitude of the line and N the RMS in the continuum) lower than 4.

\begin{figure*} 
\centering
\includegraphics[scale=0.25]{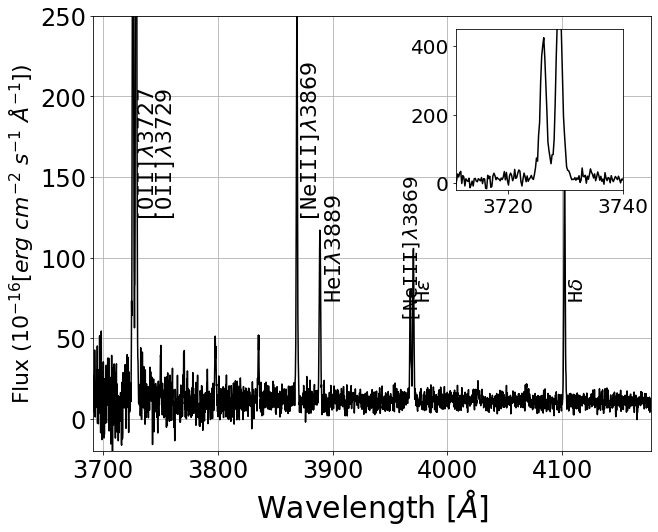}
\includegraphics[scale=0.25]{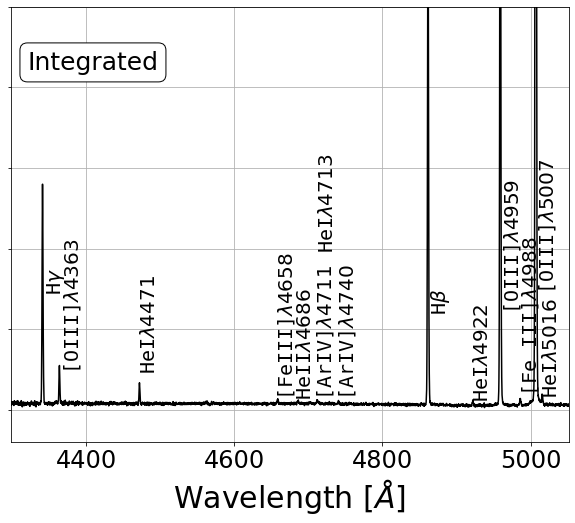}
\includegraphics[scale=0.25]{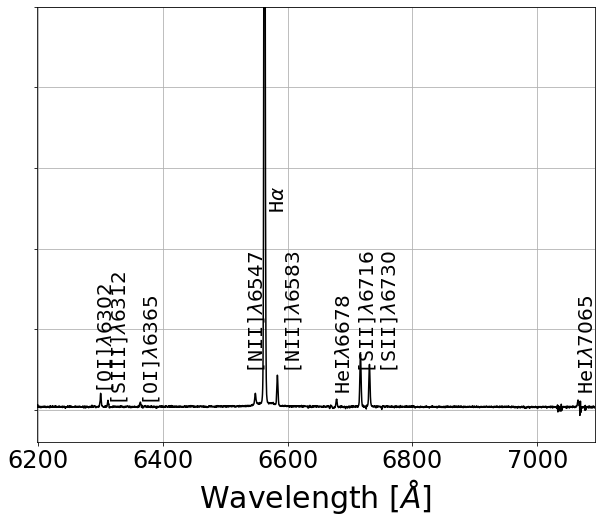}

\includegraphics[scale=0.25]{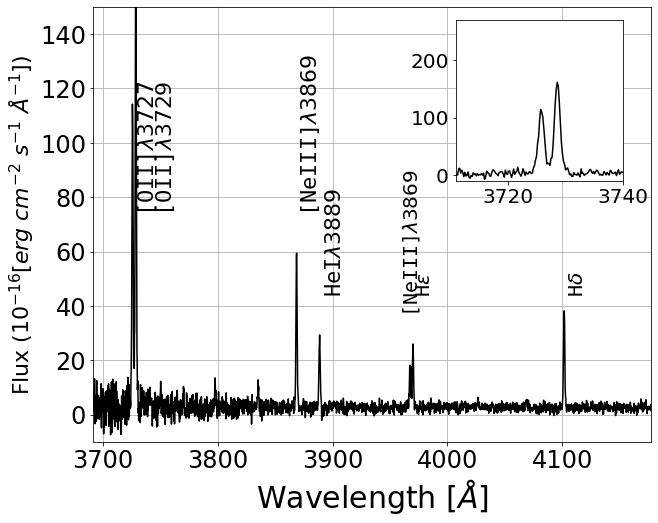}
\includegraphics[scale=0.25]{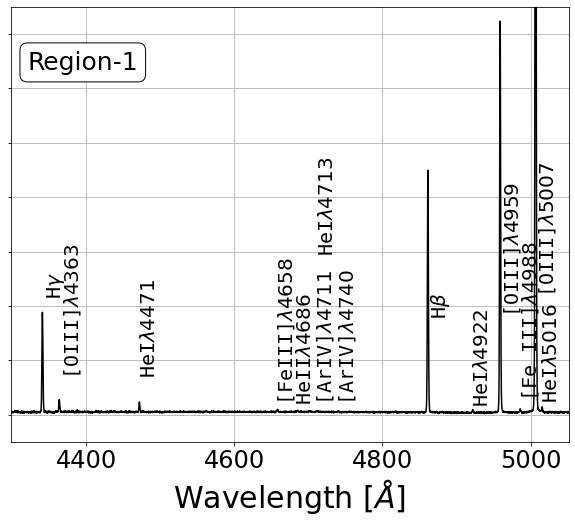}
\includegraphics[scale=0.25]{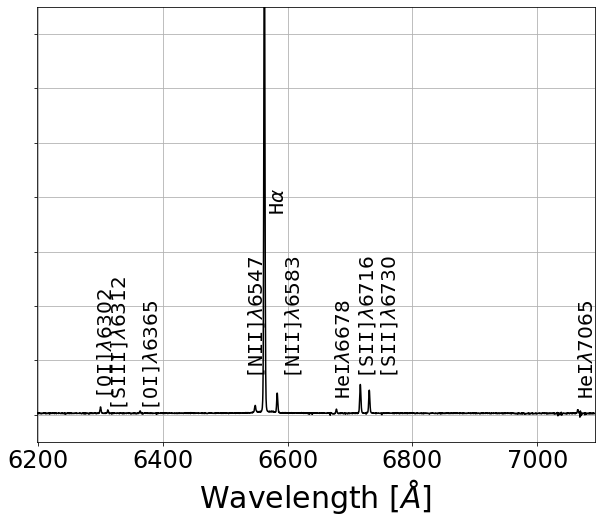}\\

\includegraphics[scale=0.25]{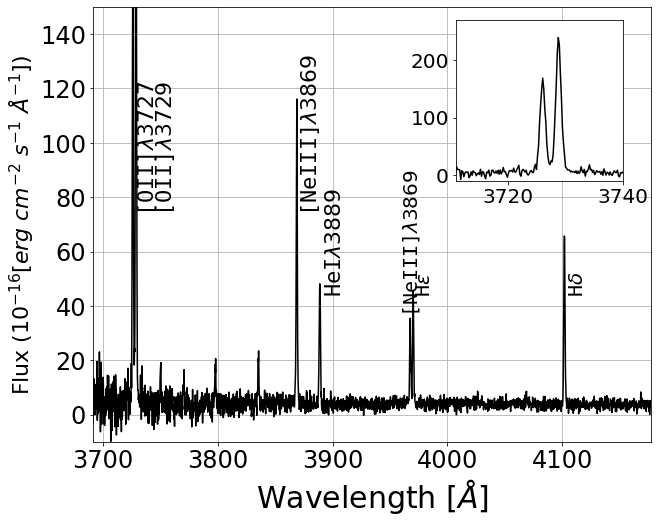}
\includegraphics[scale=0.25]{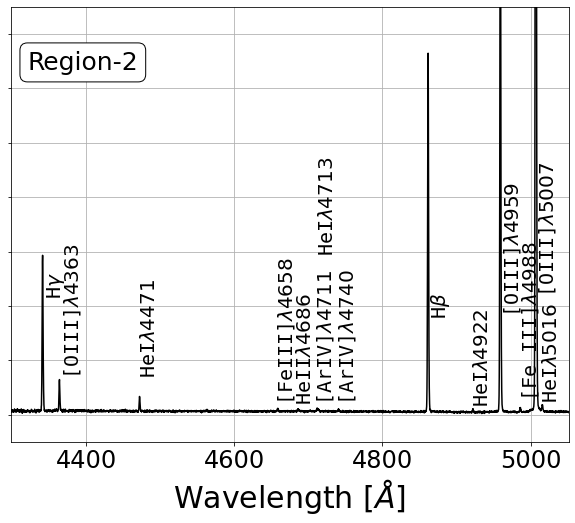}
\includegraphics[scale=0.25]{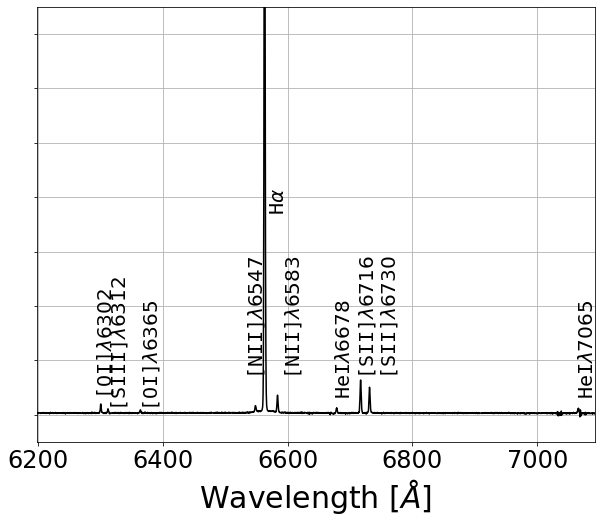}
\caption[]{Flux-calibrated spectra for J0842+1150 integrated regions. From left to right, VPH405-LR, VPH480-LR, VPH675-LR. The inset in the left column  shows  a zoom to the region of the \oii\ doublet. The label in the middle panel indicates the region where the spectrum has been obtained from.}
\label{spectra_region12}
\end{figure*}

\begin{figure*} 
\includegraphics[scale=0.2]{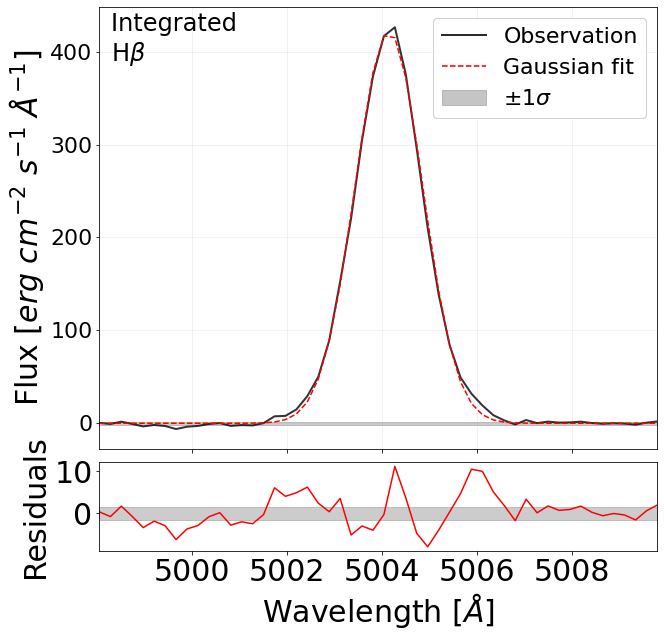}\includegraphics[scale=0.2]{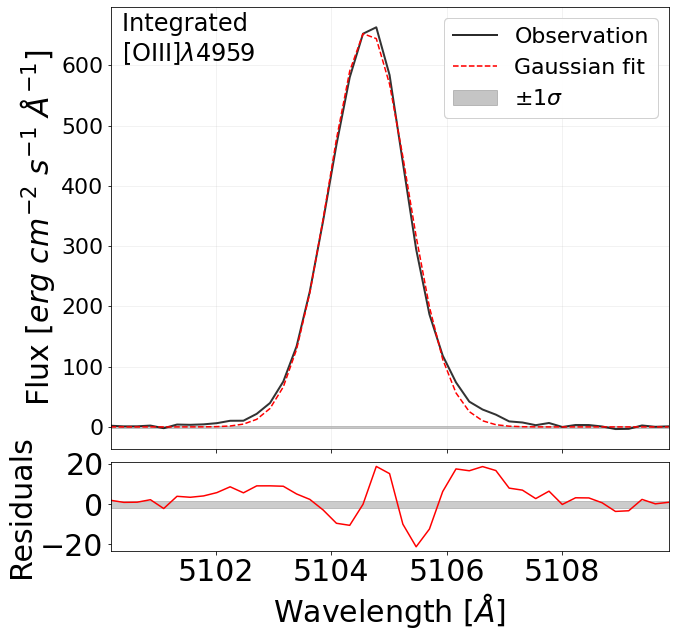}\includegraphics[scale=0.2]{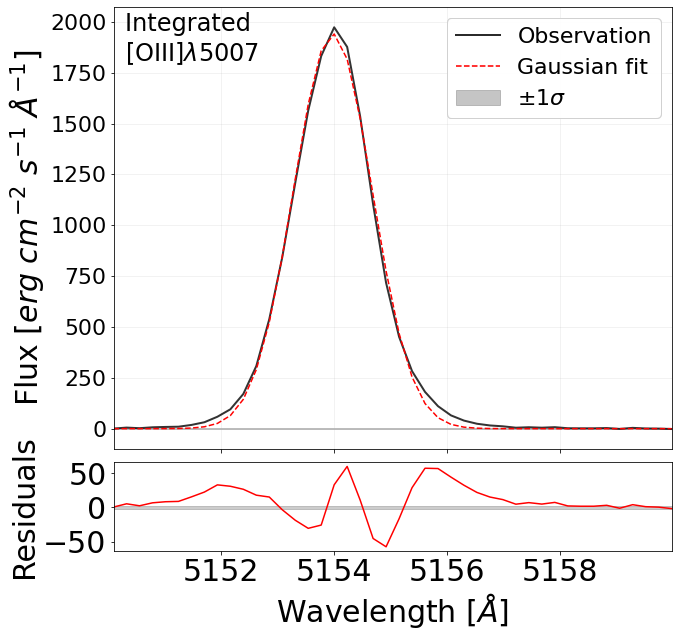}\includegraphics[scale=0.2]{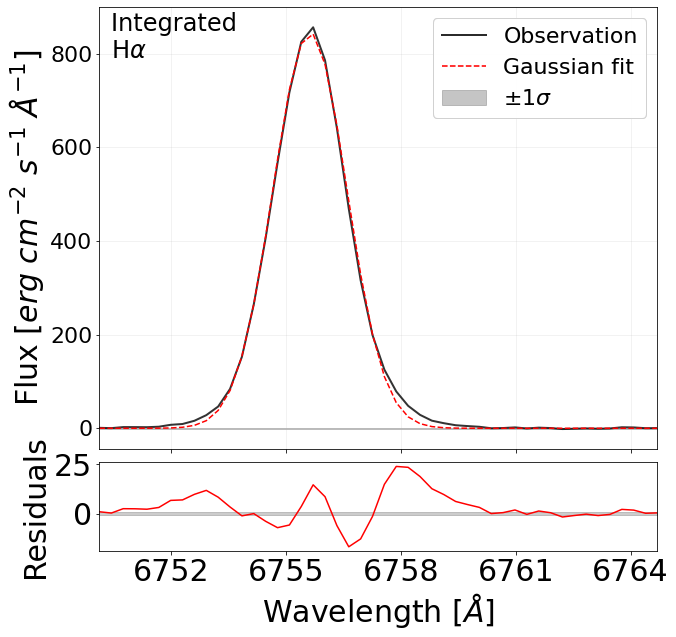}

\includegraphics[scale=0.2]{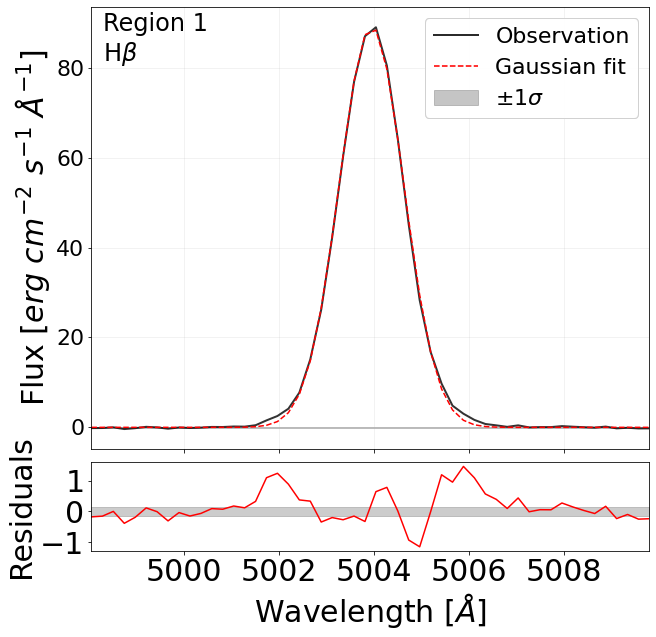}\includegraphics[scale=0.2]{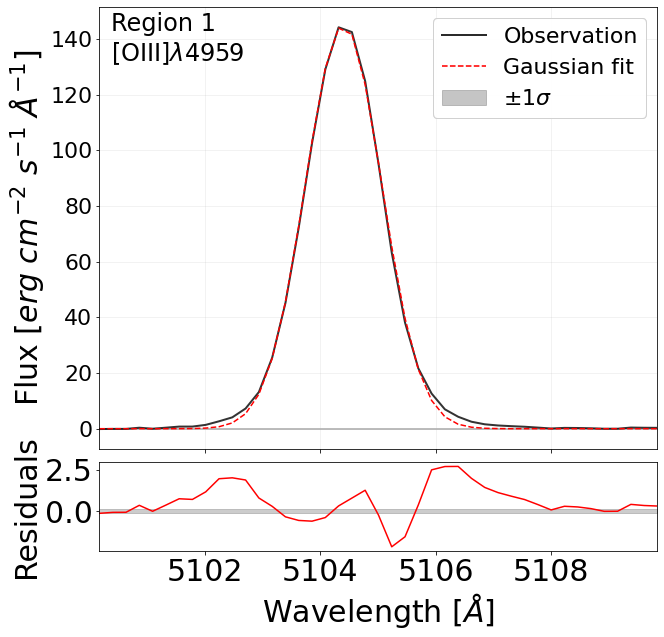}\includegraphics[scale=0.2]{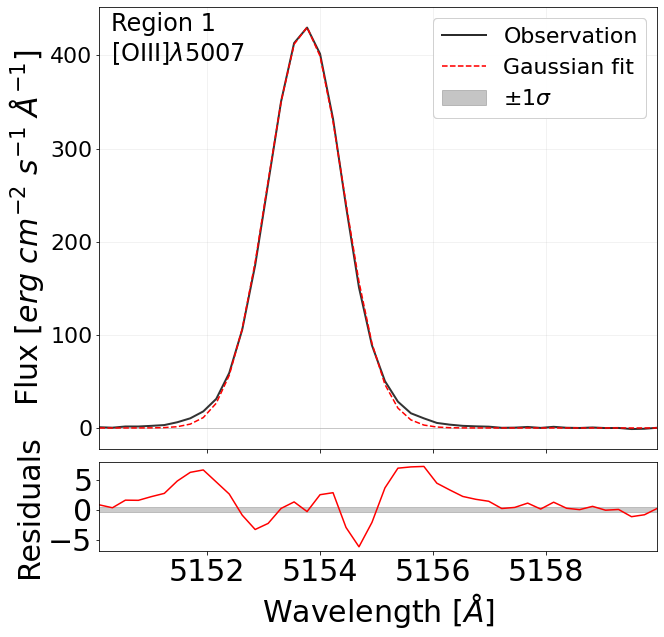}\includegraphics[scale=0.2]{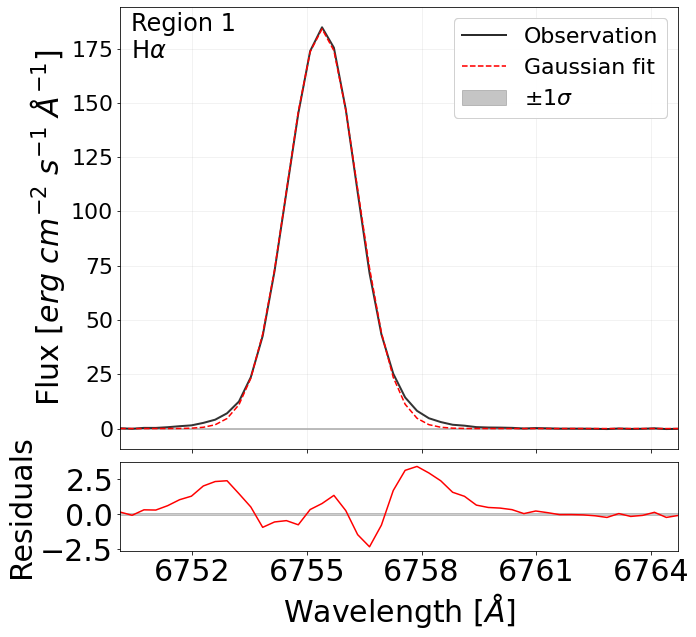}

\includegraphics[scale=0.2]{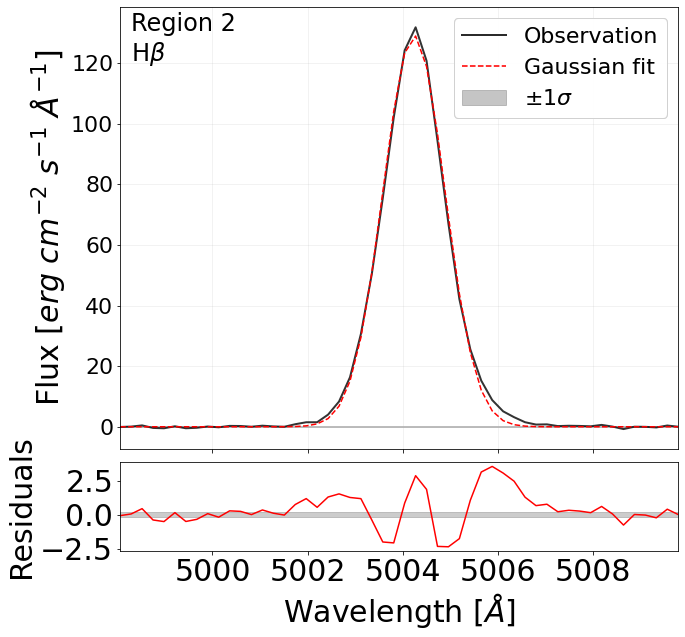}\includegraphics[scale=0.2]{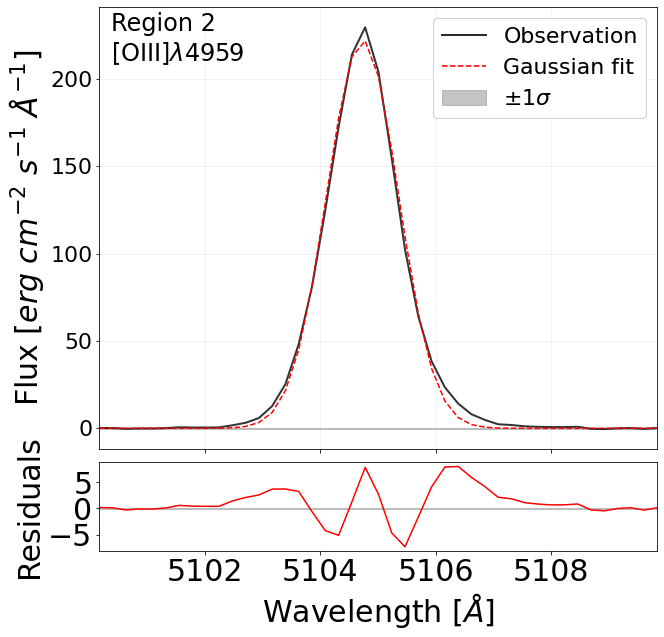}\includegraphics[scale=0.2]{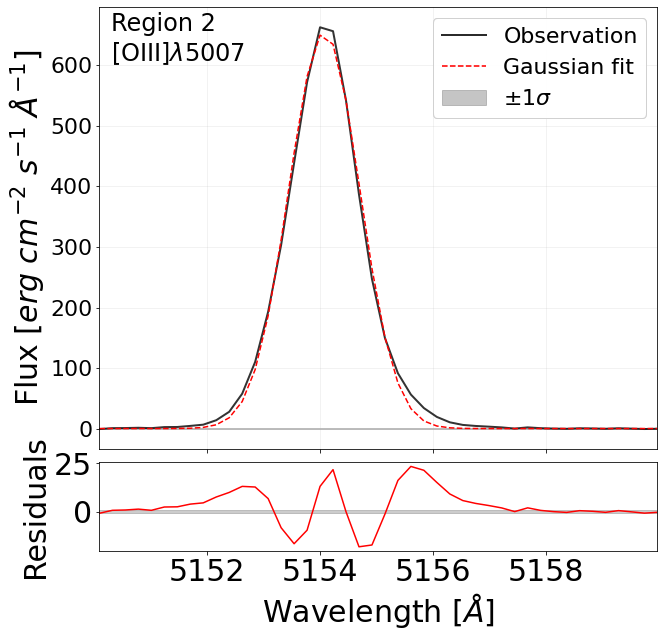}\includegraphics[scale=0.2]{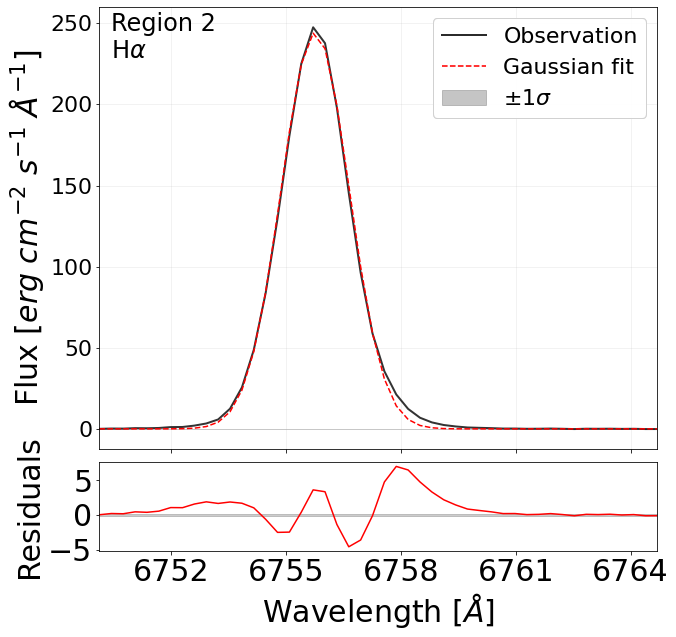}

\caption[]{Line profiles (\hbeta, \oiii$\lambda4959$, \oiii$\lambda5007$ and \halpha) and Gaussian fits for the individual regions and the integrated spectrum on the nebular emission.}
    \label{profiles}
\end{figure*}

The individual spectra for regions R1 and R2 are shown in Fig. \ref{spectra_region12}, where we also  included the integrated spectrum created by adding the flux in all the spaxels with S/N(\halpha ) (per spaxel) $\geq$4 enclosing basically all the nebular emission across the FoV. In Fig \ref{profiles}, we show individual profiles of the most intense emission lines normally found in HIIGs.

\begin{figure*} 
\centering
\includegraphics[scale=0.5]{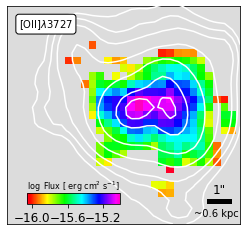}
\includegraphics[scale=0.5]{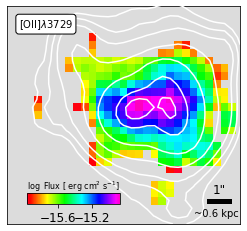}
\includegraphics[scale=0.5]{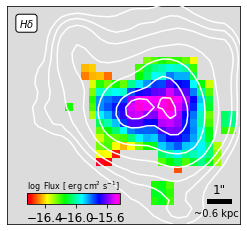}
\includegraphics[scale=0.5]{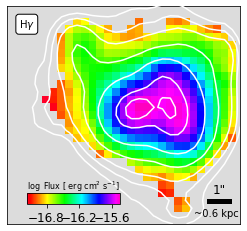}
\includegraphics[scale=0.5]{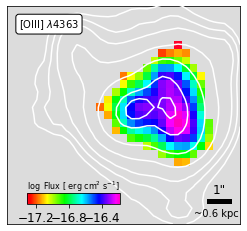}
\includegraphics[scale=0.5]{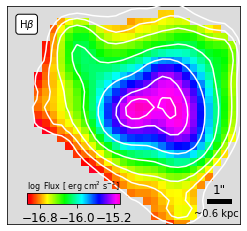}
\includegraphics[scale=0.5]{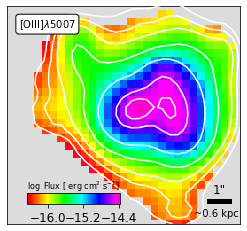}
\includegraphics[scale=0.5]{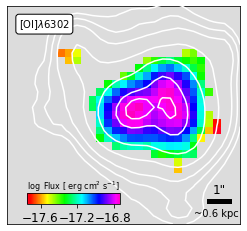}
\includegraphics[scale=0.5]{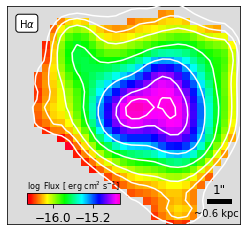}
\includegraphics[scale=0.5]{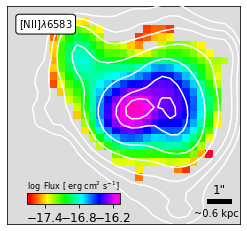}
\includegraphics[scale=0.5]{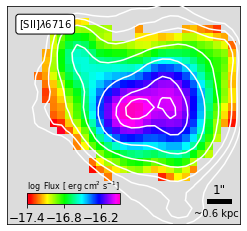}
\includegraphics[scale=0.5]{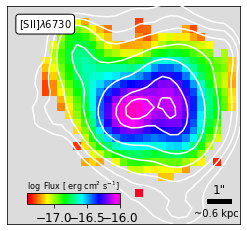}
    \caption[]{Emission line flux maps of J0842+1150: \oii$\lambda$3727, \oii$\lambda$3729, \hdelta, \hgamma, \oiii$\lambda$4363, \hbeta, \oiii$\lambda$5007, \oi$\lambda$6302, \halpha, \nii$\lambda$6583, \sii$\lambda$6716 and \sii$\lambda$6730. Only fluxes with S/N$\geq4$ are shown. The spaxels with no measurements available or below the S/N cut are left grey.  Isocontours of the \halpha\  flux are shown  for reference as in Fig. \ref{fov}.}
    \label{intensitymaps}

\end{figure*}

The intensity maps, derived from the Gaussian fits, are presented in Fig. \ref{intensitymaps}, the errors associated to the emission-line fluxes estimated by means of  the bootstrap method are between 7\% for the fainter lines (e.g. \oiii$\lambda$4363, \nii$\lambda$6583 \sii$\lambda\lambda$6716,30) and 2\% for the strong lines (e.g \halpha, \hbeta, \oiii$\lambda$5007).

\section{Spatially-resolved properties of the ionized gas}\label{properties}

\subsection{Extinction correction}

Massive bursts of star formation are embedded in large amounts of gas and dust. The latter is responsible for a wavelength-dependent light extinction along the line of sight due to absorption and scattering. Usually the amount of extinction is estimated using hydrogen recombination lines  through  the Balmer decrement, although contamination by underlying stellar Balmer absorption lines (especially for high-order ones) changes the ratio of the observed emission lines such that the internal extinction could be overestimated.

\begin{figure} 
\centering
\includegraphics[scale=0.5]{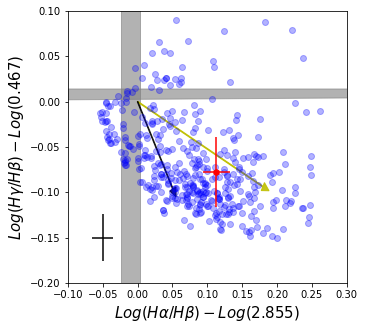}
    \caption[]{Balmer decrements for the spaxels with S/N$\geq4$ in the Balmer lines normalized to the theoretical values. The red circle marks the value derived from the SDSS spectrum. The black cross at the bottom left represents the mean error of the blue points. The vector due to pure extinction is shown in yellow; the black one shows the effect of an underlying stellar population. The vertical and horizontal bands correspond to the  Balmer decrement  theoretical values reported by \cite{Groves2012}.}
    \label{extin}
\end{figure}

To correct for extinction we used a modification of  the Balmer decrement method. We corrected the Balmer line emissions for the effect of stellar absorption lines  using the technique proposed by \cite{rosa2002}. This method allows us to obtain simultaneously the values of the  visual extinction, $A_{V}$,  and the underlying stellar absorption, $Q$. In  Fig. \ref{extin}, we show the Balmer decrement plane $\log (F(H\alpha)/F(H\beta))$ vs. $\log (F(H\gamma)/ F(H\beta))$. Extinction and underlying absorption vectors are indicated by the yellow and black arrows respectively in Fig. \ref{extin}. In the absence of underlying absorption all points should be distributed along the extinction vector, while in the absence of extinction all points should be distributed along the underlying absorption line.
 
We calculated $A_{V}$ and $Q$ using the theoretical ratios for case B recombination $F(\mathrm{H}\alpha) / F(\mathrm{H}\beta) = 2.86$ and $F(\mathrm{H}\gamma) / F(\mathrm{H}\beta) = 0.47$ \citep{ost89}. These ratios are weakly sensitive to density but have larger variations due to temperature. In Fig. \ref{extin}, we plot two gray bands corresponding to these variations due to temperatures between 5000-20000 K reported by \cite{Groves2012}; within the errors our spaxels fall within these regions. We measured the Balmer lines from the flux maps and propagated the uncertainties by a Monte Carlo procedure.  Errors in the luminosities are dominated by uncertainties in the extinction correction. In cases where the flux ratio is lower than the theoretical value we assume  a 0 value for the extinction. The dereddened fluxes for each spaxel are obtained from the expression:

\begin{equation}
 F_{0}(\lambda)=F_{\rm obs}(\lambda)10^{0.4A_{V}k(\lambda)/R_{V}}
\end{equation}
where $k(\lambda)=A(\lambda)/E(B-V)$ is given by the extinction law from \citet{Gordon2003}, and $R_{V}=A_{V}/E(B-V)$ is the optical total-to-selective extinction. We adopted  $R_{V}=2.77$ corresponding to the LMC2 supershell near the 30 Doradus starforming region, the prototypical GHIIR in the Large Magellanic Cloud. This value is the result of the LMC2 supershell sample average. Finally, the dereddened fluxes are corrected by the underlying absorption:

\begin{equation}
  F(\lambda)=\frac{F_{0}(\lambda)}{1-Q}
\end{equation}

\begin{figure} 
\includegraphics[scale=0.48]{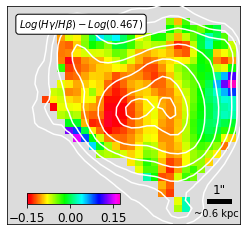}
\includegraphics[scale=0.48]{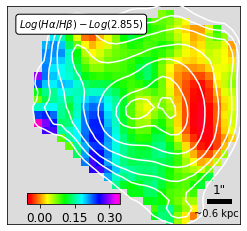}\\
\includegraphics[scale=0.48]{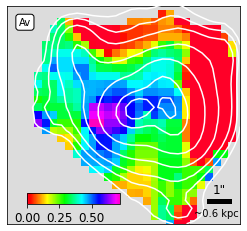}
\includegraphics[scale=0.48]{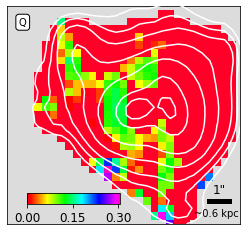}\\
\includegraphics[scale=0.48]{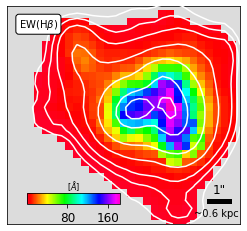}
\includegraphics[scale=0.48]{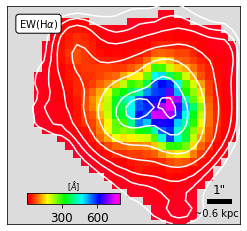}\\
    \caption[]{Upper panels: Balmer lines intensity ratio  \hgamma/\hbeta\ (left) and  \halpha/\hbeta\ (right). Middle panels: visual extinction ($A_{V}$) and underlying stellar absorption ($Q$). Bottom panels: distribution of the equivalent width of \hbeta\ and \halpha. }
    \label{extinctionmap}
\end{figure}

Fig. \ref{extinctionmap} shows the maps of equivalent width for \hbeta\ and \halpha\ and the  intensity ratios  \halpha/\hbeta\ and \hgamma/\hbeta\ normalized to their theoretical values and the maps corresponding to $A_{V}$ and $Q$.  Although this last correction is necessary, in this case it does not affect the results. The stellar continuum is very low and stellar absorptions are not detected in the spectra. 

\subsection{Diagnostic diagrams maps}

\begin{figure} 
\includegraphics[scale=0.48]{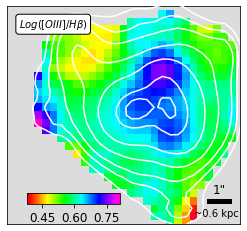}
\includegraphics[scale=0.48]{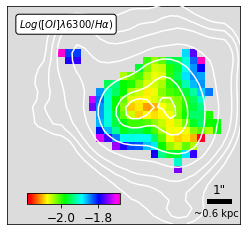}\\
\includegraphics[scale=0.48]{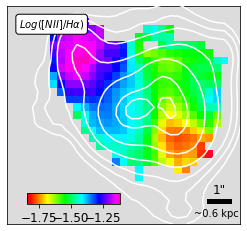}
\includegraphics[scale=0.48]{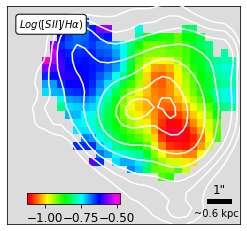}
    \caption[]{Maps of the diagnostic line ratios in logarithmic scale. The inset  indicates the line ratios  \oiii$\lambda$5007/\hbeta, \oi$\lambda$6300/\halpha, \nii$\lambda$6584/\halpha\ and \sii$\lambda\lambda$6717,31/\halpha.}
    \label{bptratio}
\end{figure}

By combining the flux maps of different emission lines, we created the spatially resolved BPT diagrams \citep{baldwin1981}, widely used to differentiate  the excitation mechanism for star-forming galaxies from  AGN. The spatial distribution for the BPT line ratios of \oiii$\lambda$5007/\hbeta, \oi$\lambda$6300/\halpha, \nii$\lambda$6584/\halpha\ and \sii$\lambda\lambda$6717,31/\halpha\ are displayed in Fig. \ref{bptratio}. The line ratios are not corrected for extinction, but this is expected to be a minor effect as the ratios involve lines which are close to each other in wavelength.

\begin{figure*} 
\centering
\includegraphics[scale=0.45]{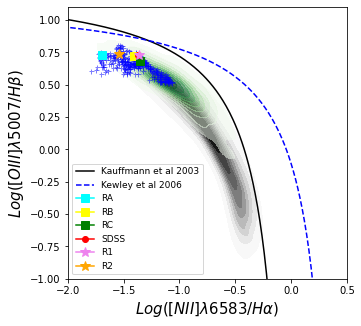}\includegraphics[scale=0.45]{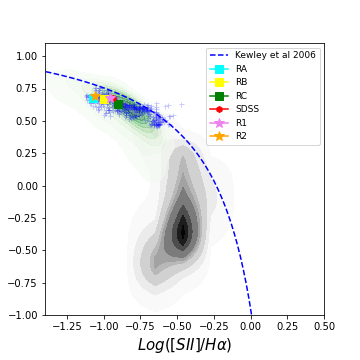}\includegraphics[scale=0.45]{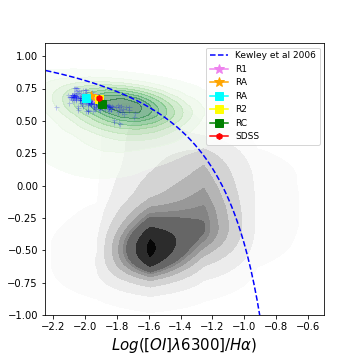}
\includegraphics[scale=0.45]{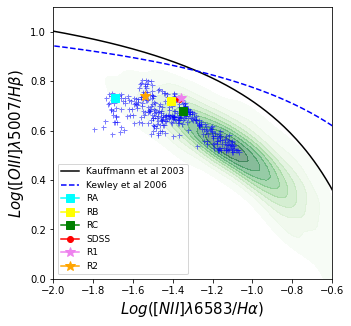}\includegraphics[scale=0.45]{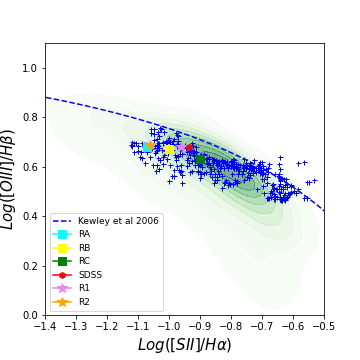}\includegraphics[scale=0.45]{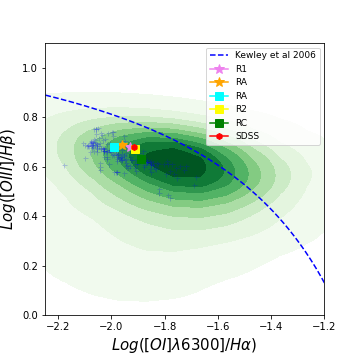}
    \caption[]{BPT diagnostic diagrams for J0842+1150. From left to right  \oiii$\lambda 5007$/\hbeta\ vs. \nii$\lambda 6583$/\halpha, \sii$\lambda 6717,31$/\halpha\ and \oi$\lambda 6300$/\halpha\ in log scale. Only fluxes with S/N$\geq$4 are shown (blue cross). The red circle marks an individual fiber from SDSS. The black  curve in the  \nii$\lambda 6584$/\halpha\ diagram represents the demarcation between star-forming galaxies (below and to the left of the curve) and AGN defined by \cite{kauffmann2003}. The blue-dashed line, in all three panels, is the theoretical demarcation limit from \cite{kewley2006}, that separates objects where the gas ionization is mainly due to hot massive stars (below and to the left of the curve) from those where other ionizing mechanism is required. The stars: violet and orange show the integrated region R1 and region R2, respectively. The squares: cyan, yellow and green represent the integrated regions labelled as RA, RB, RC respectively (see Appendix \ref{metal_regions}). Density contours come from local SDSS star-forming galaxies (DR16) separated by \eqwhbe$\sim50$\AA. Green contours for \eqwhbe$\geq$50\AA\ and black for \eqwhbe<50\AA. The bottom row shows a zoom to the region of high ionisation and low metallicity (see description in the  text).}
    \label{bptfigure}
\end{figure*}

Spatially-resolved BPT diagrams for J0842+1150 are shown in Fig. \ref{bptfigure}: \oiii$\lambda 5007$/\hbeta\ vs. \nii$\lambda 6584$/\halpha, \sii$\lambda 6717,31$/\halpha\ and \oi$\lambda 6300$/\halpha. Each point corresponds to one spaxel, with line detection threshold of $\geq4$ for each line. Fig. \ref{bptfigure} shows that for every  position in the galaxy our emission-line ratios fall in the star-forming region according to the empirical classification line of \cite{kauffmann2003} and  theoretical upper bound to pure star-formation given by \cite{kewley2006}, i.e. below and to the left of the separation lines in all BPT diagrams, specifically, in the high excitation and low metallicity region. This suggests that photoionization from hot massive stars appears to be the dominant excitation mechanism within the nebular region of J0842+1150. 

For comparison purposes, we separated into two regions the star-forming galaxies according to the equivalent width in the \hbeta\ line corresponding to the main criteria to choose samples of HIIGs. In Fig. \ref{bptfigure}, we also plot local SDSS star-forming galaxies (DR16) separated by the \eqwhbe. The green contours correspond to \eqwhbe$\geq$50\AA\ and black contours to \eqwhbe<50\AA. This value of \eqwhbe\ corresponds to an upper limit in the age of 5 Myr assuming an instantaneous starburst model, i.e. that all stars are formed simultaneously in a short-living starburst episode. For individual spaxels and integrated regions, the results are consistent with the location of the HIIGs as can be seen in the lower panels, which are a zoom to the region of high ionization and low metallicity consistent with \eqwhbe$\geq$50\AA.

The spectra of individual spaxels  occupy a narrow band on the BPT diagram consistent with the region characterized by star formation. The offset seen in the BPT diagrams of individual spaxels with respect to the SDSS HII galaxies is apparent and corresponds to the density of points that we plot for HII galaxies which have a median value of 43\AA\ in the \eqwhbe. Higher values of \eqwhbe\ tend to populate a higher region in this diagram showing a sequence in \eqwhbe\ \cite[e.g.][]{Curti2022} (or, equivalently, specific star formation rate). In our case, the \eqwhbe\ values for J0842+1150 are even greater than 50\AA\ in many spaxels. In the case of \oi$\lambda$6300/\halpha, we do not have enough spaxel \oi$\lambda$6300 detections towards low \eqwhbe\ values.

\subsection{Electron density}

The quality of the spectra  with the detection of weak emission lines, allows us to determine the $T_e$ and $n_e$ and hence the metallicity using the direct method as well as  other physical conditions of the gas. 

We have used the Python package \texttt{PyNeb}\footnote{An innovative code for analysing emission lines. \texttt{PyNeb} computes physical conditions and ionic and elemental abundances and produces both theoretical and observational diagnostic plots; we used the version 1.1.14 \url{https://github.com/Morisset/PyNeb_devel/tree/master/docs}  } \citep{Luridiana2015} to compute the physical properties and ionic oxygen abundances in a homogeneous way using the relevant line intensity ratios.

The electron density, $n_e$ and electron  temperature were calculated simultaneously from the \sii$\lambda6717$/\sii$\lambda6731$ line ratio and \oiii$\lambda$4363/\oiii$\lambda$5007  respectively. We used the atomic data and collision strengths from  \cite{Rynkun2019} and \cite{Tayal2010}, respectively given by the default dictionary in \texttt{PyNeb}. The errors have been obtained using a bootstrap method. For each spaxel, we propagated the error in the lines intensity, in the extinction correction, computed the line intensity ratios, and run \texttt{PyNeb}. This process is repeated 100 times per spaxel. We obtain a distribution of density for each spaxel and  take the median as the value of the spaxel and the standard deviation as a measure of the associated error.

\begin{figure*} 
\includegraphics[scale=0.55]{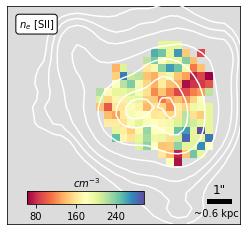}
\includegraphics[scale=0.55]{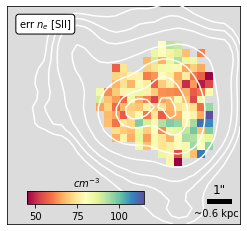}
\includegraphics[width=7cm,height=4cm]{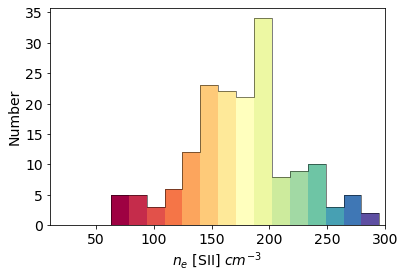}
    \caption[]{Derived electron density using the \sii$\lambda6717$/$\lambda6731$ ratio. From left to right: density map, its error  and  distribution across the nebula.  }
        \label{densitySII}
\end{figure*}

Fig. \ref{densitySII} shows the density maps, the associated error distribution maps and the histogram of density values across the galaxy derived from the \sii$\lambda6717$/\sii$\lambda6731$ line intensity ratio. We found a homogeneous density distribution from 70 to 300 cm$^{-3}$ with errors varying from 50 to 100 cm$^{-3}$ and an average value of 153$\pm$59 cm$^{-3}$. The structure observed in the density maps is not significant at more than $1\sigma$.

\begin{figure*} 
\includegraphics[scale=0.55]{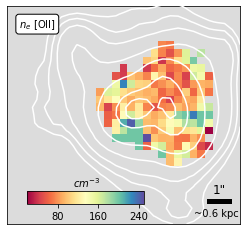}
\includegraphics[scale=0.55]{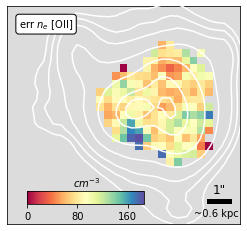}
\includegraphics[width=7cm,height=4cm]{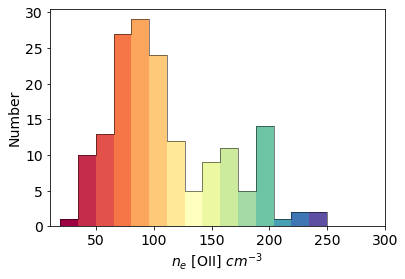}
    \caption[]{Derived electron density using \oii$\lambda3729/\lambda3726$  ratio. From left to right: density map, its error and the distribution  across the nebula.  }
            \label{densityOII}
\end{figure*}

We also computed the electron density using the ratio of \oii$\lambda3729/\lambda3726$  and the atomic data of \oii\ from \cite{Zeippen1982} and collision strengths from \cite{Kisielius2009}. The electron density distribution of the nebular emission derived from the \oii$\lambda3729/\lambda3726$  is  lower than the values derived using the sulphur lines, with a maximum of 250 cm$^{-3}$ and minimum of 30 cm$^{-3}$ with errors between 50 cm$^{-3}$ and 160 cm$^{-3}$ and an average value of 112$\pm$50. Within errors the results are consistent and do not affect the analysis and subsequent results. Fig. \ref{densityOII} shows the density derived using the \oii\ lines.

\subsection{Electron temperature}

The intensity ratios of selected forbidden lines are highly sensitive to $T_e$ and thus are normally used to calculate it. For example the ratio of \oiii$\lambda4363/(\lambda5007 + \lambda4959)$ and \nii$\lambda5755/(\lambda6459+\lambda6583)$ among others (see e.g.~\citet{hagele2007} for a comprehensive  description of the physical conditions in HIIGs), but the main disadvantage is that the auroral lines such as \oiii$\lambda4363$ or \nii$\lambda5755$ are weak and  therefore not detected if the observations are not deep enough or the metallicity of the region goes up.

The MEGARA VPHs used in this work and MEGARA resolution allow us to detect and to deblend the weak line of \oiii$\lambda4363$ across the main knots of nebular emission in the galaxy with enough S/N to obtain a reasonable estimate of the electron temperature using the ratio of  \oiii$\lambda4363/\lambda5007$. 

We used  the intensity ratio of \sii$\lambda6717$/$\lambda6731$ and \oiii$\lambda4363/\lambda5007$ to constrain the electron density and temperature at the same time, making use of the diagnostics class \texttt{getCrossTemDen} in \texttt{PyNeb}, which allows simultaneous determination of the temperature and density by fitting two different line ratios. This command returns the density and temperature, which can be used in subsequent calculations as for the determination of ionic abundances. The errors are obtained in the same way as for the electron density. 

\begin{figure}
\centering
\includegraphics[scale=0.55]{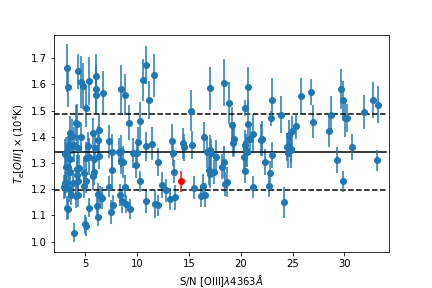}
    \caption[]{$T_e$\oiii\ derived directly from the
measurement of the \oiii$\lambda4363$ line flux vs. the relative error of the measurement. The solid horizontal line marks the mean value for $T_e$\oiii$\sim 1.33 \times 10^4$ K.  The dashed lines represent $\pm 1\sigma$. The red circle is the value derived from the SDSS spectra using the same routine of \texttt{PyNeb} and corresponds to $1.23\pm0.03 \times 10^4$ K. }
    \label{signaltonoise}
\end{figure}

Measuring the fluxes of very faint emission lines, e.g. \oiii$\lambda4363$, can lead to an overestimate  of the intensity of the line \citep[e.g.][]{Rola1994,Kehrig2004,Kehrig2016}. To check if our determination of electron temperature is  affected by our measurement  errors,  we plotted the relation between $T_e$\oiii\ and the S/N of the  \oiii$\lambda4363$ line in Fig. \ref{signaltonoise};  no systematic effects are observed. This is evidence that the largest values of $T_e$\oiii\ that we derived here are not a consequence of putative flux overestimates on the \oiii$\lambda4363$ line.

Fig. \ref{signaltonoise} displays the distribution of the T$_e$(\oiii)   which shows values varying from nearly 11000 K to 17000 K with errors between 400 K and 1000 K. These values are normally found for individual HIIGs \citep[e.g.][]{Hagele2008,Perez2009} and more recently by \cite{Perezmontero2021} where they present the distribution of more than 1000 extreme emission-line galaxies in SDSS and find a mean value of 12100 K, with a standard deviation of 1800 K.

\begin{figure*} 
\includegraphics[scale=0.55]{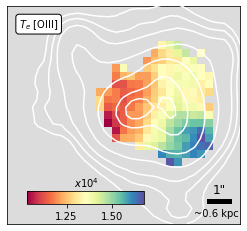}
\includegraphics[scale=0.55]{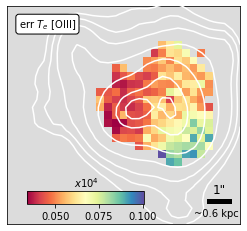}
\includegraphics[width=7cm,height=4cm]{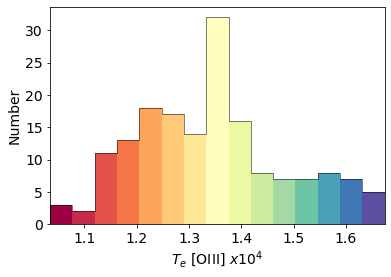}
    \caption[]{Derived electron temperature using \oiii$\lambda4363/\lambda5007$ ($T_e$\oiii) and $n_e$\sii. From left to right: $T_e$  map derived only for spaxels with fluxes in \oiii$\lambda4363\geq4$. Error in the  $T_e$\oiii\ map and distribution across the nebula. }
    \label{electronTemp}
\end{figure*}

\begin{figure*} 
\includegraphics[scale=0.55]{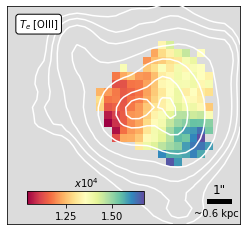}
\includegraphics[scale=0.55]{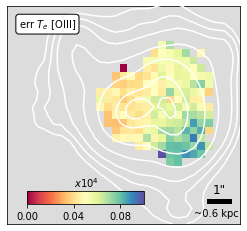}
\includegraphics[width=7cm,height=4cm]{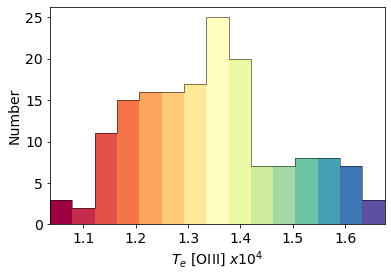}
    \caption[]{Derived electron temperature using \oiii$\lambda4363/\lambda5007$ ($T_e$\oiii) and  $n_e$\oii. From left to righ:  $T_e$ map derived only for spaxels with fluxes in \oiii$\lambda4363\geq4$, 
 error in the  $T_e$\oiii\ map and distribution  across the nebula. }
    \label{electronTempoiii}
\end{figure*}

The maps of $T_e$ with its error and the histogram of the $T_e$ distribution are shown in Fig. \ref{electronTemp}. We also compute  $n_e$\oii, from the ratio of \oii$\lambda3729/\lambda3726$, to derive $T_e$\oiii\ and the results are shown in Fig. \ref{electronTempoiii}. We did not find significant variations in the electron temperature distribution using either density indicator.

The values found here are lower than values reported for IZw18, one of the most metal-poor galaxies in the local Universe, and for which the temperature ranges between 15000~K and 22000~K \citep{Kehrig2016}, similar to the range found in SBS 0335-052E, a metal-poor blue compact dwarf galaxy, by \cite{papaderos2006} and the HIIRs in JKB 18 in the analysis by \cite{James2020}. However, the high temperatures found in these extreme galaxies are more related to their very low metallicity.

In our case, there is no evidence of higher temperatures due to shocks or other mechanism and the variations are similar to those found by \cite{Kumari2019} in the blue compact dwarf galaxy SBS 1415+437 between 11900 and 16300 K. Comparing individual galaxies with similar metallicities observed with long-slit \cite[e.g.][]{Fernandezvital2018}, the results for the individual spaxels of J084220 in terms of metallicity and temperature are consistent.

\section{Abundance derivation using the direct method}\label{abundances}

The metallicity of HIIGs is an important parameter in order to characterise their evolutionary stage and to link them to other objects that present similar properties, e.g. dwarf Irregulars (dI) or Low Surface Brightness Galaxies (LSBG). Oxygen is the most abundant of the metals in GHIIRs. Additionally, carbon and nitrogen are among the most abundant chemical elements in star-forming galaxies. Since metals are a direct product of star formation in galaxies, chemical abundances are a powerful probe of the feedback processes driving their evolution. Oxygen occupies a key role in this type of study since its gas phase abundance can be inferred from strong nebular lines easily observed in the optical wavelength range in the low redshift Universe. In this section, we derive and discuss the helium, oxygen and nitrogen abundances and the relation between them.

\subsection{Ionic helium abundance }\label{abundances_hel}

Helium is the second most common chemical species in the Universe. The study of helium abundance has the potential to unscramble the chemical evolution within galaxies. The helium abundance is frequently derived using several emission lines in the optical and NIR wavelength ranges \citep[e.g][particularly for primordial He determinations]{kunth1983,pag92,mas94,olofsson1995,Izotov2014,Fernandezvital2018,Valerdi2019}.

\begin{figure} 
\includegraphics[scale=0.48]{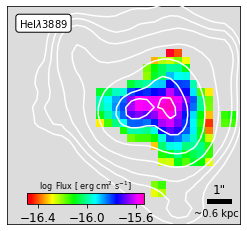}
\includegraphics[scale=0.48]{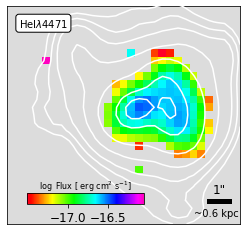}\\
\includegraphics[scale=0.48]{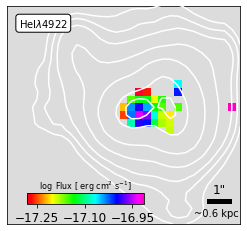}
\includegraphics[scale=0.48]{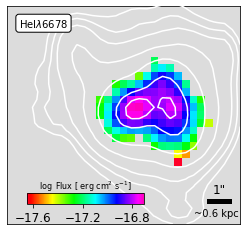}\\
    \caption[]{Maps of $\lambda$3889, $\lambda4471$, $\lambda4922$ and $\lambda4922$ for spaxels with S/N>4.  \halpha\  isocontours  are overlaid.}
    \label{helium_detection}
\end{figure}

To determine the helium abundance, we use \texttt{PyNeb} for each spaxel with detected helium lines and estimate  He$^+$/H$^+$; the adopted value is  the weighted mean. In Fig. \ref{helium_detection}, we show the maps for \hei\ $\lambda$3889, $\lambda4471$, $\lambda4922$ and $\lambda6678$. 
Given that \hei$\lambda3889$  cannot be deblended from HI H8 $\lambda3889.064$ with our spectral resolution   and \hei$\lambda4922$ is very weak and only detected in some spaxels and with large uncertainty, we used just \hei$\lambda4471$ and \hei$\lambda6678$ to compute the helium abundance assuming the derived $n_e$ and $T_e$\oiii\ for each spaxel. We use the effective recombination coefficients by \cite{Storey1995} for H and by \cite{Porter2012,Porter2013} for He which include collisional effects. The derived helium abundance for J084220 is in good agreement with values derived for HII galaxies with similar properties \cite[e.g.][]{Fernandezvital2018} and Giant HII regions in local galaxies presented by \cite{Valerdi2021} although our results present larger errors.

\begin{figure*} 
\includegraphics[scale=0.55]{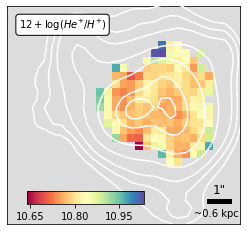}
\includegraphics[scale=0.55]{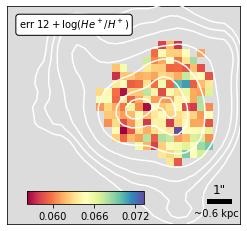}
\includegraphics[width=7cm,height=4cm]{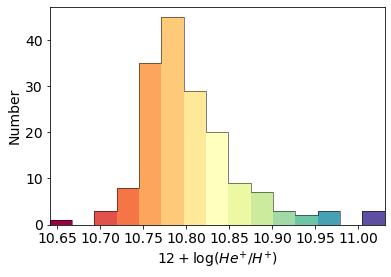}
  \caption[]{Left:  helium abundance map,  assuming $T_e$\oiii\ to quantify  He$^+$/H$^+$. Centre and right: errors map  and histogram of the distribution. The contours correspond to the \halpha\ flux.}
    \label{metallicity_helium}
\end{figure*}

Fig. \ref{metallicity_helium} shows the helium abundance (He$^+$/H$^+$) distribution. For the integrated nebular region, we found a value of \he=10.84$\pm$0.03, and for regions R1 and R2, 10.81$\pm$0.03 and 10.79$\pm$0.03 respectively. 

\subsection{Oxygen abundance}\label{abundances_oxy}

Oxygen abundance in HIIGs ranges between  $7.1\leq$\oh$\leq8.3$, obtained for more than 100 HIIGs with good quality data \citep{perez2003}. More recently \cite{Chavez2014}, using a sample of 100 HIIGs between redshifts 0.02 and 0.2, found a median value of \oh =8.08.

Here we calculate the ionic abundances of O$^+$/H$^+$ and  O$^{++}$/H$^+$ using the line intensities of \oii$\lambda\lambda3726,29$ and \oiii$\lambda\lambda4959,5007$ respectively and the corresponding electron temperature and density $T_e$\oiii\ and $n_e$\sii. These parameters can be used simultaneously in the \texttt{getIonAbundance} task in PyNeb to obtain ionic abundances.
For the low ionization region, we have used the temperature relation proposed by \cite{Esteban2009} (equation 3) to estimate $T_e$\oii\, which is valid for a temperature range between 2000 and 18000 K. Finally,  by adding the contribution of O$^+$/H$^+$ and  O$^{++}$/H$^+$  we determine the total oxygen abundances as O/H=O$^+$/H$^+$+O$^{++}$/H$^+$.

\begin{figure*} 
\includegraphics[scale=0.55]{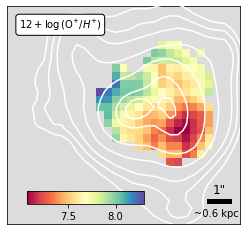}
\includegraphics[scale=0.55]{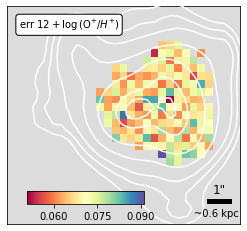}
\includegraphics[width=7cm,height=4cm]{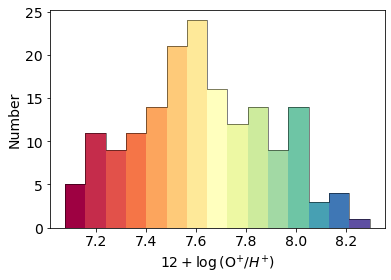}\\
\includegraphics[scale=0.55]{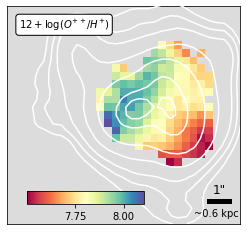}
\includegraphics[scale=0.55]{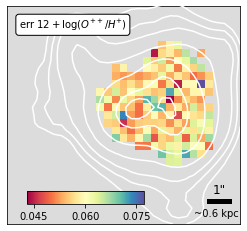}
\includegraphics[width=7cm,height=4cm]{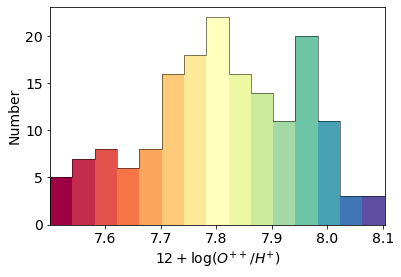}\\
\includegraphics[scale=0.55]{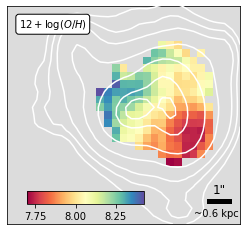}
\includegraphics[scale=0.55]{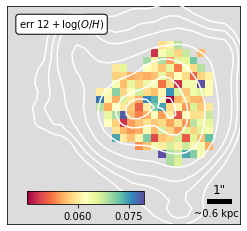}
\includegraphics[width=7cm,height=4cm]{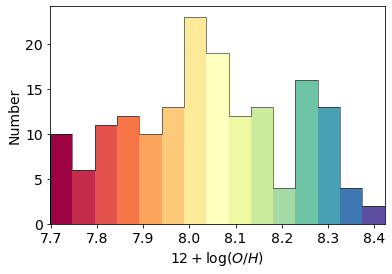}\\
    \caption[]{From top to bottom: Map of ionic and total oxygen abundance derived for \ohmas, \ohmasmas\ and \oh\ only for spaxels with derived $T_e$\oiii. Centre and  right:  corresponding errors  and histograms of the abundance distribution. }
    \label{metallicity}
\end{figure*}

The spatial distribution of the derived \ohmas, \ohmasmas\ and \oh\ are displayed in Fig. \ref{metallicity}. We also present the error map and the histogram distribution for each one. 

We find an average value of \oh=8.06$\pm$0.06, and a difference of $\Delta$O/H=0.72 dex between the minimum and maximum values (7.69$\pm$0.06 and 8.42$\pm$0.05) indicating a large metallicity range, unusual in a dwarf star-forming galaxy. A similarly  large range has been reported in a recent paper for Haro~11 \citep{Menacho2021}.

\subsection{Nitrogen abundance}\label{abundances_nitro}

In the optical, the nitrogen abundance is commonly derived from the \nii$\lambda\lambda$6548,83 lines. Our spectral resolution is enough to discriminate both lines from \halpha. We assume $T_e$(\nii)=$T_e$(\oii). We used the atomic data of \cite{FROESEFISCHER20041} and collision strengths from \cite{Tayal2011}. For all spaxels, we assumed that N/H$\approx$N$^{+}$/H$^{+}$ and thus we derive the total abundance of nitrogen. 

\begin{figure*} 
\includegraphics[scale=0.55]{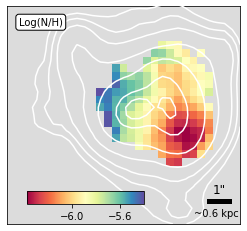}
\includegraphics[scale=0.55]{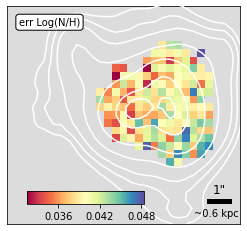}
\includegraphics[width=7cm,height=4cm]{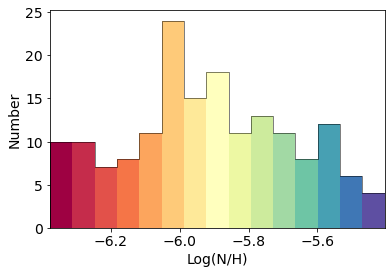}
  \caption[]{Nitrogen abundance map assuming $T_e$\nii=$T_e$\oii. Centre and right: distribution of the corresponding errors and histograms of the distribution. }
    \label{nitrogenAbundance}
\end{figure*}

Fig. \ref{nitrogenAbundance} displays the nitrogen abundance map, the associated error and the distribution across the galaxy. We found a mean value of $\log$(N/H) of -6.0, with a range between -6.4 and -5.4 and errors around 0.035 dex.

\subsection{Nitrogen to oxygen abundance ratio}

Nitrogen is mainly produced during the CNO cycle. In order to reproduce the tendency of N/O vs. O/H in star-forming galaxies, nitrogen production has been interpreted due to two main sources. The first one is short-lived massive stars that produce pure ``primary'' nitrogen and are responsible for the N/O ratio in a low metallicity plateau. The second mechanism refers to the low- and intermediate-mass stars, which produce both secondary and primary nitrogen and enrich the ISM with a time delay relative to massive stars, and cause the increase of the N/O ratio and of the metallicity. In fact  secondary nitrogen becomes dominant and the nitrogen abundance increases at a faster rate than oxygen \citep{Matteucci1986,Henry2000,Thuan2010,Wu2013,Vincenzo2016}.

\begin{figure*} 
\includegraphics[scale=0.55]{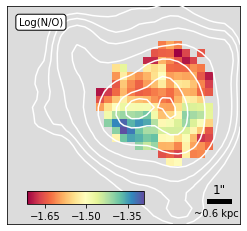}
\includegraphics[scale=0.55]{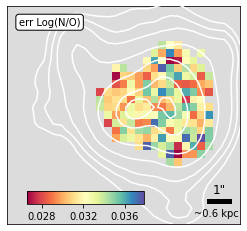}
\includegraphics[width=7cm,height=4cm]{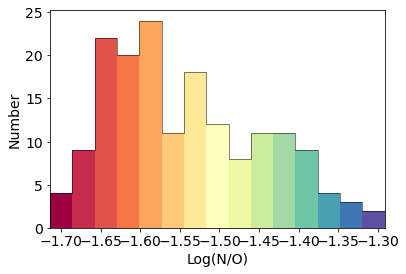}
  \caption[]{ $\log$ N/O map. Centre and right: corresponding errors and histograms of the distribution. }
    \label{metallicity_no}
\end{figure*}

Strong nebular lines in the optical range allow reliable measurement of the (N/O) abundance ratio in  star-forming galaxies, when both the \oii$\lambda\lambda3727,29$ and \nii$\lambda\lambda6548,83$ lines are available.  Using the abundances as derived in sections \ref{abundances_oxy} and \ref{abundances_nitro}, we constructed the maps and histogram of  $\log$N/O as can be seen in Fig. \ref{metallicity_no}. The  values range between $-1.7$ and $-1.30$ throughout the galaxy.

\begin{figure}
\includegraphics[scale=0.5]{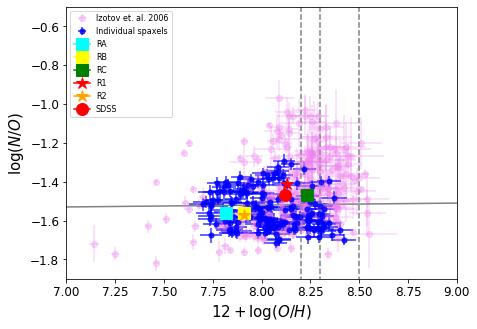}
    \caption[]{Individual spaxels in blue and integrated regions (stars and squares) in the plane  $\log$(N/O) vs. \oh. A sample of extremely metal-poor emission-line galaxies  from \cite{Izotov2006} are plotted as pink dots for comparison. The grey dashed lines represent the transition metallicity \oh\ reported by different works,  \cite{Andrews2013,Henry1999,Pilyugin2012}, corresponding to 8.5, 8.3 and 8.2 respectively. The horizontal line is the average value for all spaxels.}
    \label{NOvsO}
\end{figure}

Fig. \ref{NOvsO} displays spaxel by spaxel on the plane N/O vs O/H; for comparison, we also plotted the star-forming galaxies from \cite{Izotov2006}, with their nitrogen and oxygen abundances determined using the direct method. The average of $\log$(N/O) in this sample is $-1.45\pm$0.04. Interestingly, these galaxies have  \eqwhbe> 30\AA\ which can be considered as a criterion to choose star-forming galaxies in the zone of high ionization and low metallicity in the BPT diagram \citep{Telles2018}. 

The location of the spaxels and of the integrated regions is consistent with the location of the so called low metallicity plateau with an average \mbox{$\log$(N/O)$=-1.52\pm$0.09}, where nitrogen production is considered primary. For the integrated regions, we can also see a slight increase in the abundance of $\log$(N/O) with  increase in metallicity, but still within the low-metallicity plateau for GHIIRs in nearby galaxies and HIIGs \citep{Perez2009,Pilyugin2012,Andrews2013}. In general, we find that both for individual spaxels and for integrated regions the location of the points in the plane N/O vs. O/H correspond to primary nitrogen.


\section{Empirical Abundances}\label{abundances3} 

\subsection{Oxygen abundance}

As metallicity increases in low excitation photoionised regions,  T$_e$ cannot be determined due to the weakness (or absence) of the relevant auroral lines (e.g.
 \oiii$\lambda4363\AA $) and abundances cannot be directly derived. For these cases,  empirical methods using strong forbidden lines have been proposed and calibrated over the years \cite[e.g][]{Searle1971,Jensen1976,Pagel1978,Shields1978,Pagel1979,Alloin1979, Diaz2000a,denicolo2002,Kewley2002,Marino2013,Pettini2004}. Some of these methods (O3N2, N2) have been more recently reviewed by \cite{Marino2013,Arellano2016} and \cite{Perezmontero2021}, this latter following the direct method in the ranges \mbox{$7.7<$\oh$<8.6$}, with a sample of 1969 extreme emission-line galaxies (EELGs) at a redshift \mbox{$0\le{\rm z}\le0.49$}, selected from the SDSS. 

Nowadays, O3N2 and N2  are  the most widely used indicators of the oxygen abundance in photoionised regions. The line ratios that define these empirical parameters are given below. They are used both at low and high redshifts.  N2  is defined as the ratio between two emission lines that are so close in wavelength that the parameter is not affected by reddening and calibration effects. The lines are also accessible in the near-infrared at moderate-to-high redshifts. O3N2 is only  weakly affected by differential extinction and makes use of the strongest and most easily accessible emission lines in  rest-frame optical spectroscopy.

Additionally, the relationship between N2 and O/H is mono-valued and although it is affected by the  ionization parameter of the gas \citep[e.g.][]{denicolo2002,Marino2013}, it can be used to rank abundances in more extended regions, particularly relevant 
for integral field spectroscopy (IFS) data \citep[e.g.][]{Lopez2011,Kehrig2013}. O/H can also be derived  using the ratio between \oii\ and \oiii\ forbidden emission lines. This ratio is mostly sensitive to the ionization parameter and to the equivalent effective temperature of the ionising stars, but it can also give an estimate of the total oxygen abundance based on the relation between stellar metallicity (Z) and the ionization parameter ($\log$U) \citep{Perezmontero2021}. 

To test whether strong line methods can be used to find metallicity variations across the nebula, we compared the metallicity spaxel by spaxel derived via the direct method (based on T$_e$(\oiii$\lambda$4363/\oiii$\lambda\lambda4959,5007$) with that obtained with empirical methods  such as N2 \citep[with the calibrations by][]{Raimann2000,denicolo2002,Pettini2004,Perez2009,Perezmontero2021}, O3N2 \citep{Pettini2004,Nagao2006,Perez2009,Perezmontero2021} and O32 \citep{Perezmontero2021}. 

We use the following line ratios  to  derive the empirical oxygen abundance:

\begin{equation}\label{eq_N2}
\textrm{N2}=\log \left(\frac{\textrm{I(\nii}\lambda6583)}{\textrm{I(\halpha)}}\right)
\end{equation}

\begin{equation}\label{eq_O3N2}
\textrm{O3N2}=\log \left(\frac{\textrm{I(\oiii}\lambda5007)}{\textrm{I(\hbeta)}}\frac{\textrm{I(\halpha})}{\textrm{I(\nii}\lambda6583)} \right)
\end{equation}

\begin{equation}\label{eq_O32}
\textrm{O32}=\log \left(\frac{\textrm{I(\oiii}\lambda4959,5007)}{\textrm{I(\oii}\lambda3727)}\right)
\end{equation}

\begin{figure}
\includegraphics[scale=0.5]{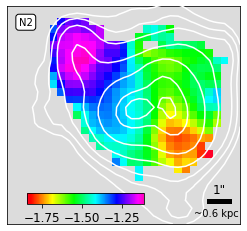}(a)\includegraphics[scale=0.5]{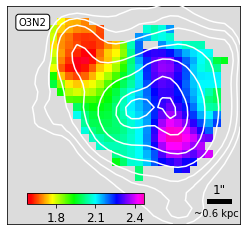}(b)
\includegraphics[scale=0.5]{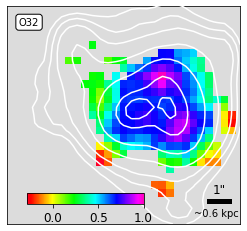}(c)\includegraphics[scale=0.5]{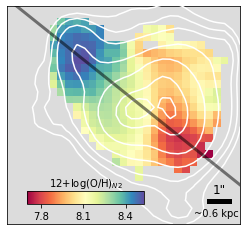}(d)
\includegraphics[scale=0.4]{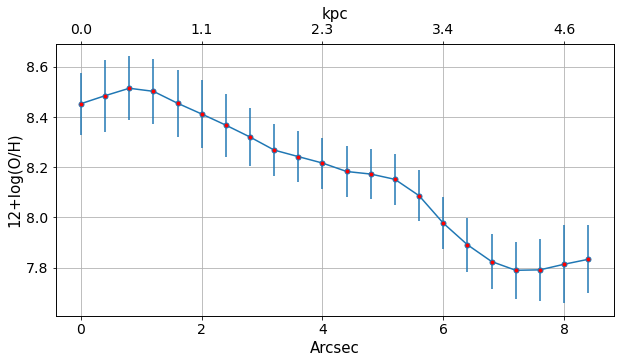}(e)
    \caption[]{(a,b,c): Maps of the line ratios N2, O3N2 and O32 defined in the text. (d) oxygen abundance map derived from the strong line method (N2); the  black line indicates the maximum metallicity variation. (e):  oxygen abundances  traced  along the line of maximum metallicity variation drawn  in  panel (d).} 
    \label{maps_calibration_oh}
\end{figure}

\begin{figure}
\centering
\includegraphics[scale=0.5]{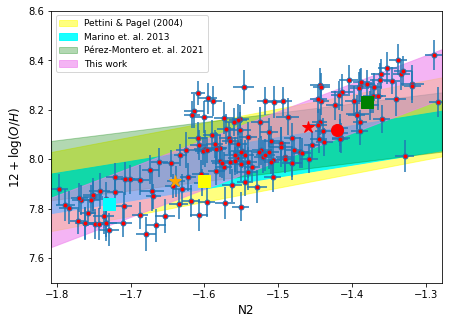}
\includegraphics[scale=0.5]{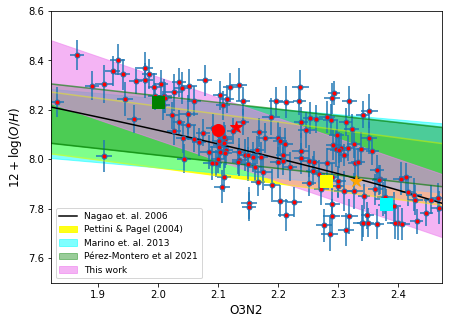}
\includegraphics[scale=0.5]{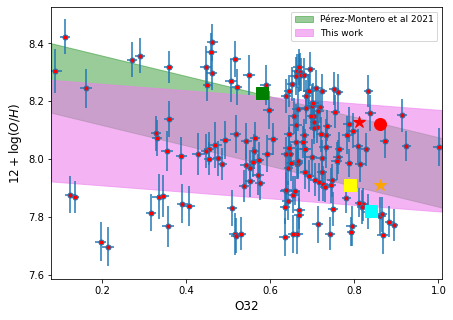}
    \caption[]{N2, O3N2 and O32 vs \oh. The symbols correspond to the integrated regions as described in Fig. \ref{bptfigure}. The coloured regions correspond to different calibrations found in the literature as described in the text. The violet region in the three plots corresponds to the linear fits using the spaxels (small red circles) and its rms; errors in both axes have been considered for the fits.}
    \label{others_indicators}
\end{figure}

These line ratios maps are shown in Fig. \ref{maps_calibration_oh}. Fig. \ref{others_indicators} shows N2, O3N2  and O32 for the spaxels with  oxygen abundances derived by the direct method.  For comparison we included \cite{Pettini2004,Marino2013,Perezmontero2021} calibrations for N2. They agree within the errors. For values greater than $-1.5$ and \oh>8 the results fall in \cite{Pettini2004} calibration. Our spaxel by spaxel results for J0842+1150 give a relation \mbox{\oh= $9.80+1.12\times {\rm N2}$} with an rms of 0.21. Following this relation it is possible  to  estimate the metallicity over a larger number of spaxels. 

For the O3N2 indicator, we also included \cite{Nagao2006} calibration, an approximation by a third degree polynomial function. For individual spaxels we find consistency between different calibrations in the range of O3N2 of $\sim-1.3$ and $-1.8$ and \oh\ between 8.1 and 8.4. Outside this range our results are more consistent with \cite{Nagao2006} calibration. 

The resulting fittings for  O3N2 and O32 are: \mbox{\oh=$9.86-0.83\times{\rm O3N2}$} and \mbox{\oh=$8.10-0.21\times{\rm O32}$} with an rms of 0.12 dex and 0.17 dex respectively. The largest discrepancies are found in  O32; consistent with \citet{Perezmontero2021} who have found that the correlation between O32 and metallicity for metal-poor galaxies is not very strong. 

In general, our results   for individual spaxels are consistent with the integrated calibrators presented in the literature. This shows that it is still possible to obtain a good determination of abundances and find gradients based on strong line methods for IFS observations where it is not possible to derive electron temperature, especially using the calibrators for N2 and O3N2 that have lower dispersion and provide the tightest correlations. 

\begin{figure}
\centering
\includegraphics[scale=0.5]{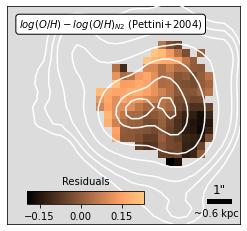}\includegraphics[scale=0.5]{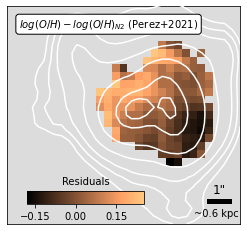}
\includegraphics[scale=0.5]{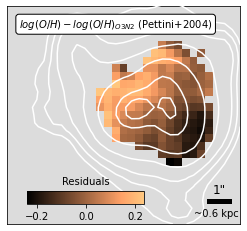}\includegraphics[scale=0.5]{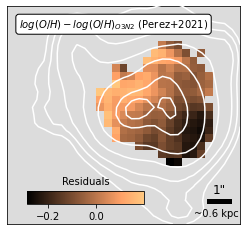}
    \caption[]{Oxygen abundance difference between the  direct method and strong-line methods. This variation is around 0.2 dex in the N2 calibrator and 0.25 dex for the O3N2 in the regions of high and low temperatures. The variations are similar whether using the N2 or O3N2 calibrators from \cite{Pettini2004} or \cite{Perezmontero2021}.}
    \label{metodh}
\end{figure}

Fig. \ref{metodh} shows the differences in the oxygen abundances derived from the direct method, which has a tight dependence on the temperature, and the strong-line method which uses strong line ratios only. In most spaxels in between the bursts of star formation, we derive consistent values. However, we find differences of 0.2 dex in the results of individual spaxels for ratios at low and high N2 and O3N2, corresponding to the most extreme temperatures found in the galaxy around $\sim$11000 K and $\sim$16000 K.
 
The strong line methods are secondary calibrated methods that could be degenerate with respect to some properties of the nebula like pressure and ionization parameter, therefore give imprecise values in the high and low abundance ends and fail to probe the diverse conditions in one galaxy, especially for IFU observations. Nevertheless they still represent a powerful tool to estimate the abundances in absence of weak lines to measure temperatures \citep{Dopita2013}. Although we can not find a direct explanation for those differences, we suggest that part of these discrepancies could be originated from a temperature  effect  and that additional bias could be related to the atomic data used in the empirical derivations, compared with the set used here \cite[e.g][]{Perezmontero2021} Similar results and conclusions have been found by \cite{Menacho2021}.

\subsection{Strong-line methods and the nitrogen-to-oxygen abundance ratio}\label{abundances4}

N/O can be derived using a combination of strong collisional  lines in the form of  the N2O2 parameter  \cite{Perez2009}. A problem though is that the lines used have a significant wavelength difference which renders the parameter more sensitive to observational issues, like calibration errors or the fact that a single observing  setup may not yield all the relevant lines  simultaneously. Therefore, \cite{Perez2009} also suggest the use of  N2S2 to characterize the abundance ratio between nitrogen and oxygen. These indicators are given by: 

\begin{equation}\label{n2o2}
\textrm{N2O2}=\log \left(\frac{\textrm{I(\nii}\lambda6583)}{\textrm{I(\oii}\lambda3727)}\right)
\end{equation}

\begin{equation}\label{n2s2}
\textrm{N2S2}=\log \left(\frac{\textrm{I(\nii}\lambda6583)}{\textrm{I(\sii}\lambda6717,6731)}\right)
\end{equation}

\begin{figure}
\centering
\includegraphics[scale=0.5]{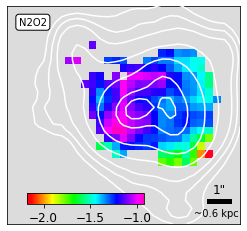}\includegraphics[scale=0.5]{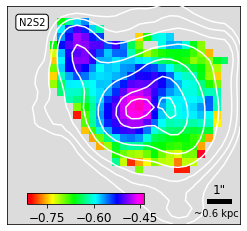}
    \caption[]{N2O2 (left) and N2S2 (right) maps.  \halpha\  isocontours  are overlaid.}
    \label{maps_indicators_n2o2_n2s2}
\end{figure}

Maps of N2O2 and N2S2 are shown in Fig. \ref{maps_indicators_n2o2_n2s2}. It can be noticed that N2S2 reaches farther than N2O2, towards the east of the galaxy.

\begin{figure}
\centering
\includegraphics[scale=0.5]{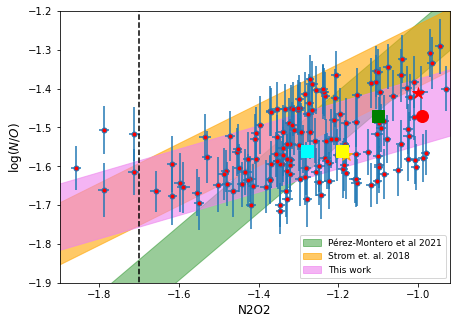}
\includegraphics[scale=0.5]{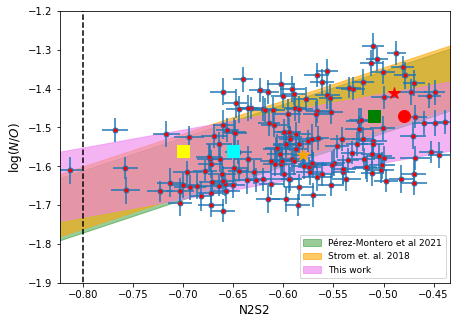}
    \caption[]{Relation between $\log$(N/O), as derived following the direct method, and the parameters N2O2 and N2S2 for the individual spaxels and for the integrated regions; the vertical dashed lines show the lowest value for the calibrations to be valid. The region in green corresponds to \cite{Perezmontero2021} calibration whereas the region in yellow is derived from \cite{Strom2018} using high-z star-forming galaxies;  in violet is shown the fit for the spaxels in this work. The large solid  points represent the different integrated regions as described in the text.}
    \label{indicators_n2o2_n2s2}
\end{figure}

A tight correlation was found by \cite{Perezmontero2021} 
between the above calibrators and N/O. The relations,  given by: \mbox{$\log{\rm(N/O)}=-0.31+0.87\times{\rm N2O2}$} and \mbox{$\log{\rm (N/O)}=-1.005+0.857\times{\rm N2S2}$} are valid in the ranges \mbox{$ -1.7<{\rm N2O2}<-0.5$},  and  \mbox{$-0.8<{\rm N2S2}<0.3$}, which correspond to  \mbox{$-1.8<\log{\rm (N/O)}<-0.75$} and \mbox{$-1.7<\log{\rm (N/O)}<-0.75$}, with standard deviation of 0.04 dex and 0.08 dex, respectively. These calibrations  are plotted in Fig. \ref{indicators_n2o2_n2s2}, where we also show our relation between these parameters and the nitrogen-to-oxygen ratio as derived via the direct method both for the spaxels across the galaxy and for the chosen integrated regions. 

Our data spaxels fit best  \cite{Perezmontero2021} calibration  for  N2S2. For  N2O2  the points depart from \cite{Perezmontero2021} fit. Our  fits  are: \mbox{$\log{\rm (N/O)}=-1.15+0.30\times{\rm N2O2}$} and \mbox{$\log{\rm (N/O)}=-1.15+0.64\times{\rm N2S2}$} with an rms of 0.089 dex and 0.085 dex, respectively. 

Our spaxel by spaxel N2O2 values agree better with \cite{Strom2018} calibration for N/O derived for high-z starforming galaxies than with \cite{Perezmontero2021}.  For  N2S2 vs $\log$(N/O), the agreement of this work and derived local and high-z calibration by \cite{Perezmontero2021} and \cite{Strom2018} respectively are consistent within the errors.

\section{Chemical inhomogeneities in HIIGs and BCD}\label{Oinhomogeneities}

The oxygen abundance of the ionized gas in HIIGs has been found to be homogeneously distributed on spatial scales of $>$100pc \citep[e.g.][]{Lagos2009,Cairos2009,Perez2009,Garcia2012,Kehrig2016}. It has to be said however, that the \halpha\ distribution in some of the  galaxies in these works,  does not show multiple star-forming knots. In general, only weak gradients are observed on scales of hundreds of parsecs, e.g.  IIZw70 in \cite{Kehrig2008} or Tol 0104-388 and Tol 2146-391 in \cite{Lagos2012}.

The apparent homogeneous distribution of abundances in HIIGs may be due to the fact that the newly synthesized elements were rapidly dispersed  on time scales <10$^7$ yr and mixed in the ISM.  Processes associated with massive star formation are extremely efficient at mixing hot and warm ionized gas. This hypothesis requires that the metals be ejected and mixed homogeneously within the ISM during the HII region lifetime to ensure the mixing of individual clouds \citep{Roy1995,Kobulnicky1997}.

A possible explanation for the homogeneous chemical appearance of HIIGs on spatial scales of >100 pc was proposed by \cite{ten96} in which the ejecta from stellar winds and SNe could undergo a long ($\sim$100 Myr) cycle in a hot phase ($\sim$10$^6$ K) before mixing with the surrounding ISM. Thus, the new metals processed and injected by the current star formation episode are possibly not observed and remain for longer in the hot gas phase, whereas the metals from previous events are well mixed and homogeneously distributed through the whole extent of the galaxy. 

On the other hand, \cite{Lagos2009} for UM408 and \cite{Izotov2006} in SBS 0335-052E, using IFU observations  find a variation of $\sim$0.5 dex in both galaxies and a common point   is that the maximum oxygen abundance found  does not correspond to the position of any of the identified bursts of star-formation. Similar variation is found in NGC 6822, a dwarf irregular galaxy in the local group, with an average of \oh$\sim8.15$  \citep{Lee2006}.

\cite{James2020} present a summary of other dwarf galaxies, Haro 11,  UM 448, NGC 4449, NGC 5253 and Mrk 996 among others, highlighting that chemical maps of (O/H) show differences of $\sim$0.5dex. They conclude  that the chemical inhomogeneities should be the result of outflows of metal-enriched gas due to SNe, the accretion of metal-poor gas due to interactions/mergers, self-enrichment from winds of massive stars, or bursts of star-formation at shorter times compared with the timescale of metal mixing.

Finally, as discussed in \cite{Bresolin2019}, there are a handful of dwarf-irregular galaxies that actually have well-ordered gradients that are comparable to spiral galaxies. For example, NGC 6822, NGC 4449, DDO 68 each show a chemical abundance gradient, despite the absence of spiral structures, deemed necessary for gradients. This links the existence of these gradients to recent enhancements in their star-formation activity, where metal mixing timescales are longer than the time between bursts of star-formation.

Analysing metal-poor galaxies, \cite{SanchezAlmeida2013,SanchezAlmeida2014,SanchezAlmeida2015} and \cite{OlmoGarcia2017} concluded that the observed metallicity inhomogeneities and gradients are only possible if the metal-poor gas fell onto the disk recently.  Among several possibilities for the origin of the metal-poor gas, they favour the infall predicted by numerical models. Thus, if this interpretation is correct, metal-poor galaxies trace the cosmic web gas in their surroundings. Their results are consistent with assuming that the local galaxies characterized by a bright peripheral clump on a faint tail, are discs in the early stages of assembling, their star formation being sustained by accretion of external metal-poor gas.

The low metallicity found for J0842+1150 (20\% of solar) is consistent with it being a metal-poor galaxy. Normally, these galaxies have been morphologically identified as cometary or tadpoles, and this association between low metallicity and tadpole or cometary shapes suggests that those are attributes characteristic of very young systems \citep{SanchezAlmeida2013}. A possible formation scenario has been proposed for  which the impacting gas clouds have the largest effect on the outskirts of galaxies where the ambient pressure and column density are low which,  for most orientations of the galaxy, results in a cometary shape \citep{SanchezAlmeida2015}.

The existence of chemical inhomogeneities is particularly revealing because the timescale for mixing in disc galaxies is short, of the order of a fraction of the rotational period. This implies that the metal-poor gas in such galaxies was recently accreted from a nearly pristine cloud, very much in line with the expected cosmic cold-flow accretion predicted to build disc galaxies \citep[][and references therein]{OlmoGarcia2017}. However, the fate of the metals released by massive stars in \hii regions is still an open question, and the processes of metal dispersal and mixing are very difficult to model.

\begin{figure}
\centering
\includegraphics[scale=0.4]{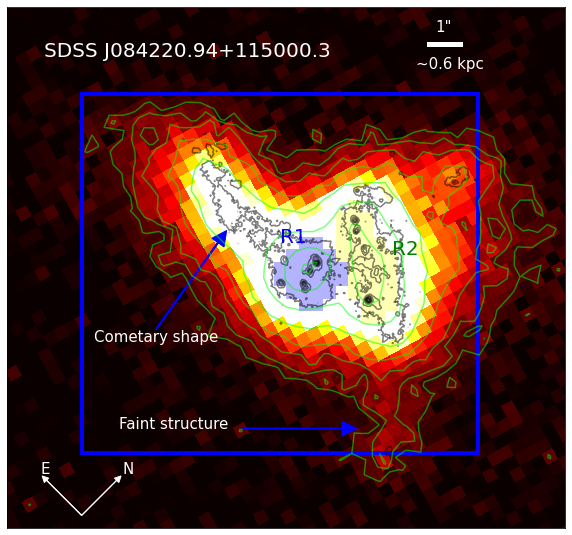}
    \caption[]{Broad-band SDSS image in \gb-band. The main star-forming regions are indicated:  region R1 in blue, region R2 in yellow, a very faint tail feature can be identified towards the south-west of region R2 and a cometary shape to the east of region R1.  Isocontours of \gb-band from SDSS are shown in green, whereas in black are shown isocontours of the WFC3 UVIS2 F336W from HST between $4.55\times10^{-20}$ and $1.5\times10^{-18}$ \uflux\AA$^{-1}$ corresponding to the minimum and maximum, respectively. The blue square is the FoV of MEGARA.}
     \label{compositeimage}
\end{figure}

In Fig. \ref{compositeimage}, we show the  image in \gb-band from SDSS of J0482+1150, where the extremes of metal content are presented; the lowest metallicity region with a metal content of about 1/10th of Solar and the eastern extension, a linear structure, with the highest value of metallicity in the galaxy (using N2 calibrator), about half the Solar value. The high value of the metal content of the north-east structure and its shape suggest an outflow of enriched material.

A possible interpretation for the  marked difference in oxygen abundance and starburst activity that we find could be that  we are witnessing an ongoing interaction system triggering multiple star-forming regions localized in two dominant clumps of different metallicity. From the distribution of the equivalent width of Balmer lines (akin to age) the more symmetric cluster (R1) shows lower EW(\hbeta) with an underlying tail or a cometary shape towards the east, whereas the more elongated one (R2) shows larger EW(\hbeta) and also the lowest metallicity.
 
Additionally, a very faint tail feature can be identified towards the south-west of region R2. Kinematic analysis of this data will allow us to look for signs of interaction in the rotation patterns. Briefly, the velocity field shows an apparent rotational motion with maximum velocity of 60 \kms, with a position angle of the kinematic axis, $\theta_{kin}=124\pm9$, as determined by a tilted-ring analysis of the emission lines velocity field based on a kinemetry method developed by \cite{Krajnovic2006}. However, these rotation motions look disturbed, perhaps as a consequence of multiplicity of star forming regions or merger systems. Towards region \#1, the velocity field appears blue-shifted with velocity between $-5$ and $-15$\kms\ whereas region \#2, more elongated, shows two velocity components: one about 15 \kms and another between $30-35$ \kms. The zero velocity is defined in the intermediate region between regions \#1 and \#2. A detailed analysis will be presented in a forthcoming paper (Fern\'andez-Arenas et. al. in prep).

\begin{table*}
\caption{Emission line-fluxes and physical properties from selected regions and SDSS fiber.}
\label{tableIntegrated12}
\begin{tabular}{|l|l|l|l|l|}
\hline \hline 
Wavelenght (\AA) & Region 1 (R1) & Region 2 (R2) & Integrated & SDSS Fiber \\ 
\hline 
3727 \oii	&	147.6	$\pm$	5.0	&	190.3	$\pm$	5.1	&	905.3	$\pm$	54.2	&	162.0	$\pm$	2.6	\\
3729 \oii	&	203.8	$\pm$	4.9	&	274.2	$\pm$	5.1	&	1345.3	$\pm$	56.1	&	143.7	$\pm$	2.4	\\
3869 \neiii	&	72.7	$\pm$	1.6	&	123.3	$\pm$	2.5	&	464.6	$\pm$	18.4	&	72.4	$\pm$	1.0	\\
3889 \hi+\hei	&	39.0	$\pm$	1.7	&	66.7	$\pm$	1.9	&	258.2	$\pm$	19.5	&	37.1	$\pm$	0.9	\\
3968 \neiii	&	18.5	$\pm$	1.5	&	38.2	$\pm$	1.9	&	136.1	$\pm$	17.9	&	50.1	$\pm$	1.2*	\\
3970 \hepsi	&	30.4	$\pm$	1.5	&	50.5	$\pm$	1.9	&	189.2	$\pm$	16.0	&	50.1	$\pm$	1.2*	\\
4102 \hdelta	&	51.0	$\pm$	1.2	&	82.9	$\pm$	1.7	&	306.3	$\pm$	13.3	&	48.5	$\pm$	1.5	\\
4341 \hgamma	&	61.4	$\pm$	0.7	&	86.9	$\pm$	1.8	&	321.4	$\pm$	3.0	&	81.2	$\pm$	1.0	\\
4363 \oiii	&	7.4	$\pm$	0.3	&	15.7	$\pm$	0.5	&	47.1	$\pm$	2.9	&	9.1	$\pm$	0.4	\\
4471 \hei	&	5.4	$\pm$	0.2	&	7.6	$\pm$	0.3	&	29.5	$\pm$	2.9	&	8.1	$\pm$	0.5	\\
4658 \feiii	&	1.7	$\pm$	0.2	&	1.8	$\pm$	0.3	&	7.1	$\pm$	2.1	&	3.4	$\pm$	0.2	\\
4686 \heii	&	0.8	$\pm$	0.2	&	1.6	$\pm$	0.3	&	3.4	$\pm$	1.7	&	-	 		\\
4711 \ariv	&	1.0	$\pm$	0.3	&	1.5	$\pm$	0.3	&	5.6	$\pm$	2.9	&	-	 		\\
4713 \hei	&	-	&	1.0	$\pm$	0.3	&	5.6	$\pm$	2.9	&	-	 		\\
4740 \ariv	&	0.7	$\pm$	0.2	&	1.0	$\pm$	0.2	&	7.9	$\pm$	4.6	&	-	 		\\
4861 \hbeta	&	151.2	$\pm$	0.7	&	209.7	$\pm$	1.2	&	744.2	$\pm$	6.7	&	185.4	$\pm$	1.5	\\
4922 \hei	&	1.6	$\pm$	0.2	&	1.8	$\pm$	0.3	&	9.9	$\pm$	3.4	&	0.8 $\pm$0.2	 		\\
4959 \oiii	&	239.0$\pm$	1.1	&	343.7	$\pm$	2.3	&	1114.3	$\pm$	12.1	&	294.2	$\pm$	2.3	\\
4988 \feiii	&	2.6$\pm$	0.2&		3.0$\pm$0.21	&	12.2$\pm$2.0	&	1.2	$\pm$	0.3	\\
5007 \oiii	&	712.7$\pm$	4.2	&	1027.0	$\pm$	7.6	&	3333.9	$\pm$	34.9	&	885.2	$\pm$	6.0	\\
5016 \hei	&	2.5	$\pm$	0.3	&	3.5	$\pm$	0.2	&	14.6	$\pm$	2.8	&	-	 		\\
6302 \oi	&	4.9	$\pm$	0.2	&	5.9	$\pm$	0.2	&	25.6	$\pm$	1.8	&	8.4	$\pm$	2.3	\\
6312 \siii	&	2.3	$\pm$	0.2	&	2.8	$\pm$	0.2	&	6.7	$\pm$	1.4	&	5.4	$\pm$	0.6	\\
6364 \oi	&	2.0	$\pm$	0.2	&	2.3	$\pm$	0.2	&	11.7	$\pm$	1.9	&	3.1	$\pm$	0.4	\\
6547 \nii	&	5.4	$\pm$	1.3	&	4.7	$\pm$	1.9	&	21.9	$\pm$	9.0	&	-	 		\\
6563 \halpha	&	422.4	$\pm$	1.3	&	544.1	$\pm$	1.9	&	2045.5	$\pm$	8.4	&	686.6	$\pm$	5.2	\\
6583 \nii	&	14.6	$\pm$	1.3	&	12.5	$\pm$	1.9	&	63.1	$\pm$	8.1	&	26.2	$\pm$	0.3	\\
6678 \hei	&	3.6	$\pm$	0.2	&	4.2	$\pm$	0.2	&	17.9	$\pm$	2.0	&	-	 		\\
6716 \sii	&	25.6	$\pm$	0.2	&	26.9	$\pm$	0.3	&	133.4	$\pm$	2.4	&	47.0	$\pm$	0.6	\\
6730 \sii	&	19.8	$\pm$	0.2	&	20.4	$\pm$	0.3	&	103.1	$\pm$	2.3	&	32.9	$\pm$	0.5	\\
7065 \hei	&	3.6	$\pm$	0.7	&	4.7	$\pm$	0.8	&	18.4	$\pm$	8.1	&	8.7	$\pm$	0.6	\\
\hline
$A_v$        & 0.601$\pm$0.15 & 0.452$\pm$0.12 & 0.521$\pm$0.11 & 0.673$\pm$0.16 \\ 
$Q\times100$ & 0.02           & 0.0 & 0.0 & 0.01 \\ 
\lum(\ulum)  & 40.73$\pm$0.14 & 40.78$\pm$0.11    & 41.23$\pm$0.12    & 40.85$\pm$0.13 \\
\eqwhbe      & 172$\pm$6 & 214$\pm$8 & 102$\pm$4 & 117$\pm$5 \\ 
\eqwhal      & 458$\pm$11 & 592$\pm$9 & 238$\pm$5 & 656$\pm$11 \\ 
$n_e$(\sii) (cm$^{-3}$)        & 153$\pm$59    & 159$\pm$85  & 161$\pm$54  & 152$\pm$38  \\ 
$T_e$(\oiii)$\times10^{4}$ (K) & 1.25$\pm$0.05 & 1.41$\pm$0.06 & 1.54$\pm$0.08 & 1.23$\pm$0.04 \\ 
\ohmas    & 7.71$\pm$ 0.06  & 7.49$\pm$0.07 & 7.61$\pm$0.07 & 7.66$\pm$0.06 \\ 
\ohmasmas & 7.91	$\pm$ 0.05  & 7.77$\pm$0.06 & 7.82$\pm$0.06 & 7.94$\pm$0.05 \\ 
\oh       & 8.13$\pm$ 0.06  & 7.91$\pm$0.06 & 8.03$\pm$0.06 & 8.12$\pm$0.05 \\ 
$\log$(N/O) & $-1.41\pm0.05$ & $-1.57\pm0.05$ & $-1.51\pm0.05$ & $-1.47\pm0.05$ \\ 
\he       & 10.81$\pm$ 0.03  & 10.79$\pm$0.03 & 10.84$\pm$0.03 & 10.84$\pm$0.03 \\ 
\hline 
\multicolumn{5}{c}{Velocity dispersion measurements (\kms)}   \\
\hline 
$\sigma$(\hbeta)	             & 35.3$\pm$1.5 & 33.9$\pm$1.5 & 37.9$\pm$1.4  &\\
$\sigma$(\oiii$\lambda4959$)	 & 33.2$\pm$1.2 & 30.2$\pm$1.3& 34.9 $\pm$1.1  &\\
$\sigma$(\oiii$\lambda5007$)	 & 33.4$\pm$1.4 & 30.5$\pm$1.3& 35.1 $\pm$1.3  &\\
$\sigma$(\halpha)	         & 35.7$\pm$1.1 & 34.3$\pm$1.1& 38.0 $\pm$1.3  &\\
\hline 
\multicolumn{5}{@{}p{11cm}@{}}{Fluxes are in units of $10^{-16}$ \uflux. $^*$  In the SDSS spectrum the emission lines  \neiii  $\lambda$3969 and \hepsi ($\lambda$3970) are blended. Here, we report the sum of the two lines for the SDSS. The velocity dispersion was obtained from the FWHM of the Gaussian fit to the profile of the emission lines and corrected by thermal, instrumental  and fine structure ($\sigma_{\rm fs}$ for \hbeta\ and \halpha\ lines only)  broadening.}\\
\end{tabular}
\end{table*}

\section{Summary  and conclusions}\label{summaryandconclu}

In this work we present MEGARA IFU spectroscopy of the metal-poor galaxy J0842+1150. We derived the distribution of physical properties across the FoV of MEGARA and integrated properties for the galaxy as a whole and for individual star-forming regions. 

The resolution, spatial coverage and high quality of data of MEGARA IFU allowed  the detection in a large number of spaxels of the auroral forbidden emission line \oiii$\lambda$4363 with a signal-to-noise level S/N $>$ 4$\sigma$.   The spaxels with detection of \oiii$\lambda$4363 include the main star-forming regions and their  surrounding area allowing the mapping of the metal distribution in J0842+1150 using the direct method.

\begin{itemize}

\item[--] Our main result is the detection of an unusually large metallicity range in a dwarf star-forming galaxy. The distribution of oxygen abundance, computed using the direct method, has a mean value of \oh=8.03$\pm$0.06 and a range of $\Delta$(O/H) = 0.72 dex between the minimum 7.69$\pm$0.06 and maximum  8.42$\pm$0.05 values. This range implies a  metallicity span  of a factor of 5 over a spatial extent of $\sim$1 kpc.

\item[--] We find that for both individual spaxels and integrated regions the location of the points in the diagram N/O vs. O/H corresponds to primary nitrogen with similar location to metal-poor galaxies reported in the literature.

\item[--] The analysis of the emission line ratios and BPT diagrams discard the presence of any AGN activity or shocks as the ionization source in this galaxy.

\item[--] The comparison of the strong-line methods to derive oxygen and nitrogen abundances for individual spaxels and integrated region are in general in agreement whithin the errors  with the calibrations reported in the literature specially the N2 and O3N2 tracers of oxygen abundance and N2S2 for the ratio of N/O. This supports the use of strong-lines methods and calibrations for spatially resolved IFS data where in many cases the detection of weak emission lines is not possible.

\item[--] The integrated values derived in this work for the physical properties of J0842+1150, such as abundances and global velocity dispersion are in agreement with values previously derived in the literature with long-slit or fiber spectroscopy.

\item[--] Among the possible mechanisms to explain the starburst activity and the large variation of oxygen abundance in this galaxy, our data support a possible scenario where we are witnessing an ongoing interaction of very young systems. A detailed multi-component kinematic analysis will be presented in a forthcoming paper (Fern\'andez-Arenas et. al. in preparation).
\end{itemize}

The study of HIIGs can offer detailed insight into processes that might depend on metallicity and that could have played an important role in the early evolution of galaxies. We plan to extend this project to objects similar to J08422+1150 (selected from \cite{Chavez2014}). We hope to use also 3D spectroscopy which will allow us to map in detail their interstellar medium chemical abundances and possible inhomogeneities and to link them to other properties like the ionising stellar fabric and morphology and their kinematics.

\section*{Acknowledgments}

We are grateful to the referee for a very thorough report that helped us improve the quality of the paper. DFA work is funded by a Consejo Nacional de Ciencia y Tecnolog\'ia  (CONACyT, Mexico) grant through project A1-S-22784. This publication is based on data obtained with the MEGARA instrument at the GTC, installed in the Spanish Observatorio del Roque de los Muchachos, in the island of La Palma. MEGARA has been built by a Consortium led by the Universidad Complutense de Madrid (Spain) and that also includes the Instituto de Astrof\'\i sica, Optica y Electr\'onica (INAOE, Mexico), Instituto de Astrof\'\i sica de Andaluc\'\i a (CSIC, Spain), and the Universidad Polit\'ecnica de Madrid (Spain). MEGARA is funded by the Consortium institutions, GRANTECAN S.A. and European Regional Development Funds (ERDF), through Programa Operativo Canarias FEDER 2014-2020.  YDM thanks CONACYT for the research grant CB-A1-S-25070 and DRG for the research grant CB-A1-S-22784 from which the postdoctoral grant that supported DFA was obtained. RC also thanks CONACyT for the research grant CF-320152. RA acknowledges support from ANID Fondecyt Regular 1202007. ALGM acknowledges support from the Spanish Ministry of Science and Innovation, project PID2019-107408GB-C43 (ESTALLIDOS), and from Gobierno de Canarias through EU FEDER funding, project PID2020010050. JMA acknowledges the support of the Viera y Clavijo Senior program funded by ACIISI and ULL. JIP acknowledges finantial support from projects Estallidos6 AYA2016-79724-C4 (Spanish Ministerio de Economia y Competitividad), Estallidos7 PID2019-107408GB-C44 (Spanish Ministerio de Ciencia e Innovacion), grant P18-FR-2664 (Junta de Andalucía), and grant SEV-2017-0709 ``Center of Excellence Severo Ochoa Program'' (State Agency for Research of the Spanish MCIU).

\section{DATA AVAILABILITY}

The data underlying this article will be shared on reasonable request to the corresponding author.

\bibliographystyle{mnras}
\bibliography{biblio_mega}

\appendix

\section{Appendix: Metallicity selected regions}\label{metal_regions}

\begin{figure}
\centering
\includegraphics[scale=0.35]{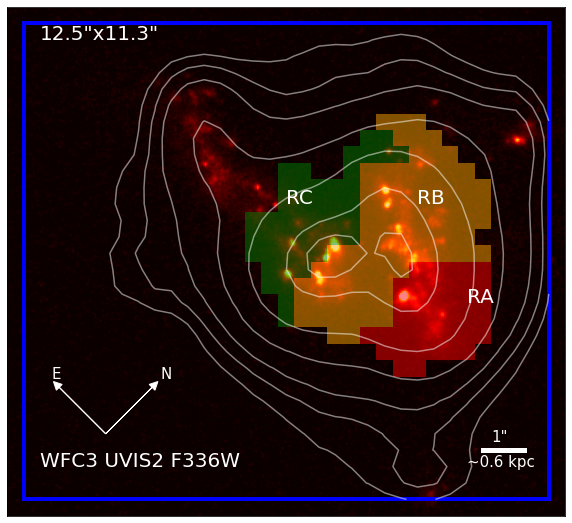}
    \caption[]{Masked regions for the extracted spectra. These have been identified according to the metallicity range as  described in the text, \oh:  <7.9 (RA),   7.9 <\oh< 8.2 (RB) and \oh>8.3 (RC). \halpha\ flux isocontours are shown in white. }
    \label{regionmetal}
\end{figure}

We noted that the dispersion of the metallicity distribution is larger than the  individual errors. In order to check this, we separated the nebula into three regions  in metallicity \oh:  <7.9,   7.9 <\oh< 8.2 and \oh>8.3. We combined the spaxels in each range in order to obtain individual spectra of these regions. The selected regions are shown in Fig. \ref{regionmetal}. 

\begin{figure*} 
\includegraphics[scale=0.26]{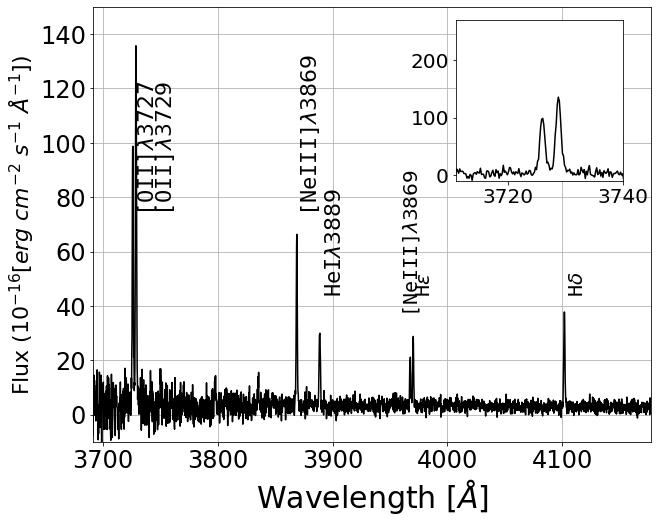}
\includegraphics[scale=0.26]{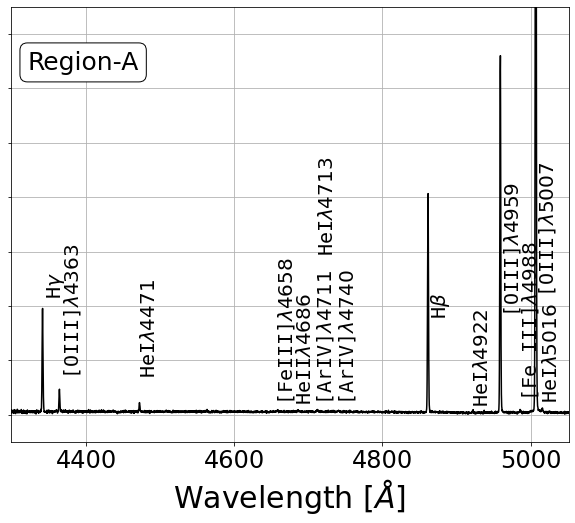}
\includegraphics[scale=0.26]{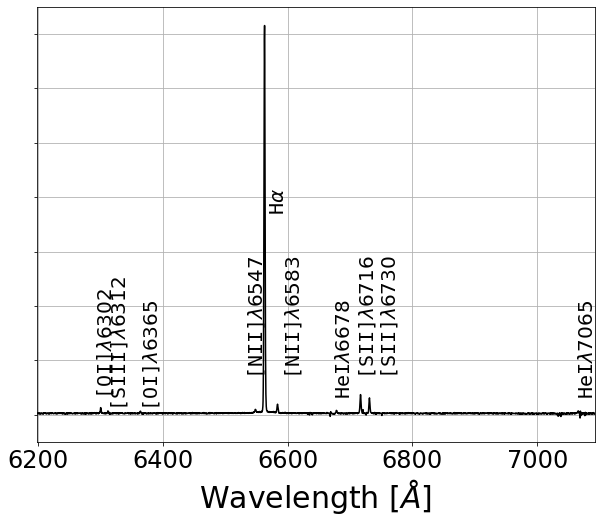}\\
\includegraphics[scale=0.26]{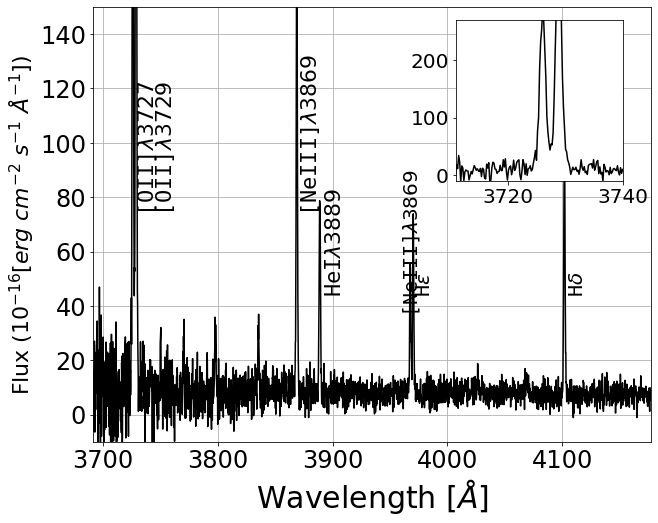}
\includegraphics[scale=0.26]{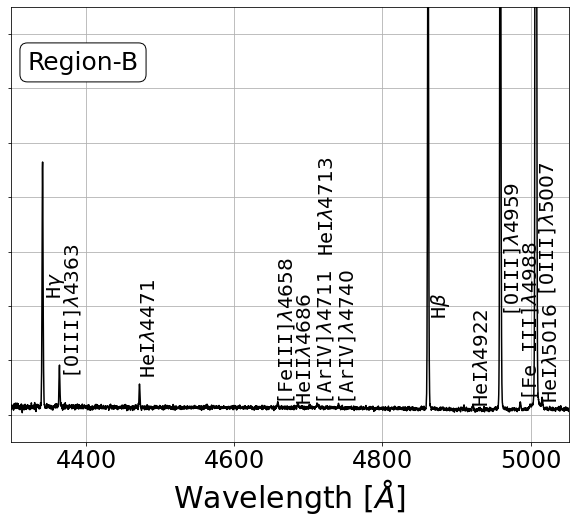}
\includegraphics[scale=0.26]{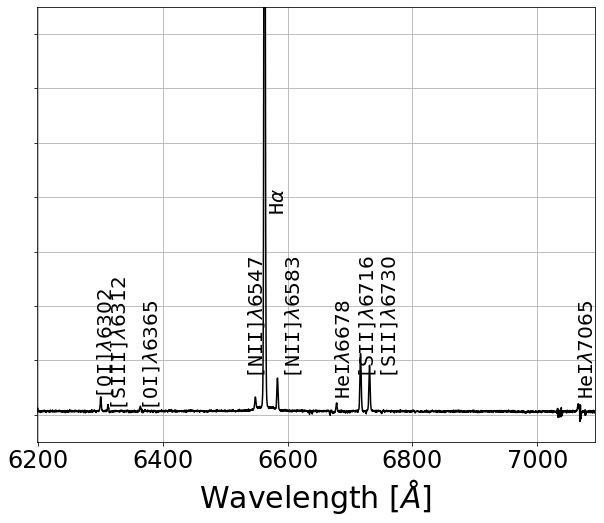}\\
\includegraphics[scale=0.26]{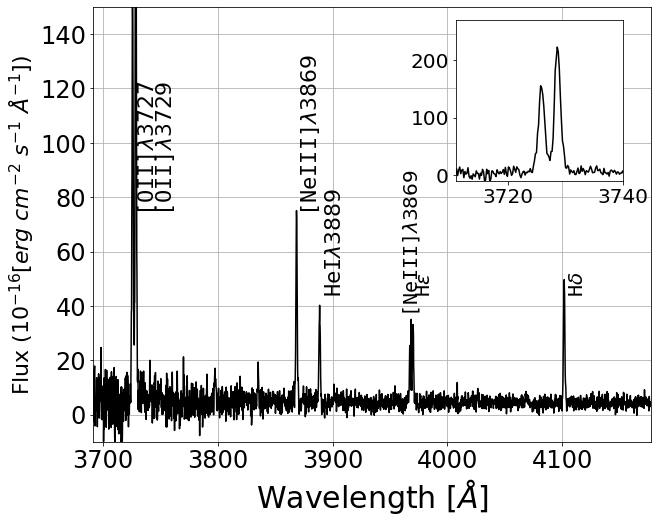}
\includegraphics[scale=0.26]{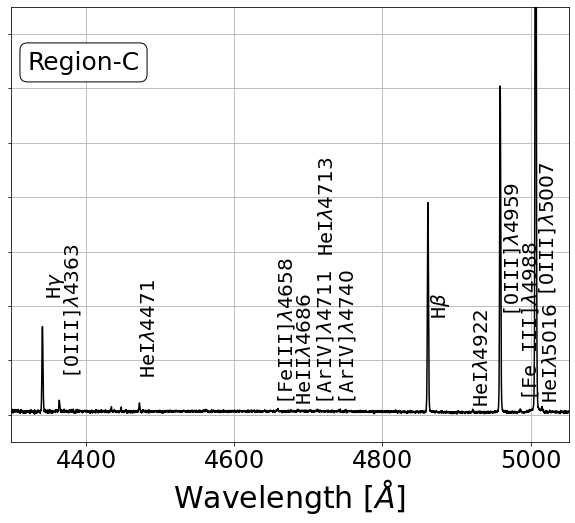}
\includegraphics[scale=0.26]{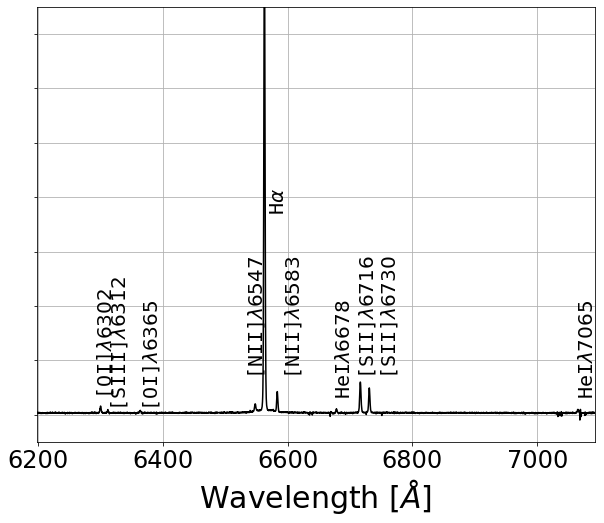}
    \caption[]{Individual spectra extracted according to the metallicity range as  described in the text. The inset in the middle panels indicates the selected regions.}
    \label{regionmetalspec}
\end{figure*}

The properties derived for each region are summarized in Table \ref{tablemetalregion} and the Fig. \ref{regionmetalspec} shows the respective spectra labelled  A, B and C. We will investigate in a forthcoming companion paper possible correlations between the metallicity inhomogeneities, the age of the individual regions as monitored by the equivalent width of \hbeta\ and the kinematics.

\begin{table}
\caption{Emission line-fluxes and physical properties from  regions selected according to metallicity.}
\label{tablemetalregion}
\begin{tabular}{|l|l|l|l|}
\hline \hline 
Wavelenght (\AA) & Region A  & Region B  & Region C  \\ 
&  (RA) &  (RB) &  (RC) \\ 
\hline 
3727 \oii	&	110.7	$\pm$	4.9	&	361.4	$\pm$	12.1	&	206.1	$\pm$	7.7	\\
3729 \oii	&	152.1	$\pm$	5.0	&	518.9	$\pm$	12.1	&	287.3	$\pm$	7.5	\\
3869 \neiii	&	69.6	$\pm$	2.1	&	206.4	$\pm$	4.7	&	96.7	$\pm$	3.1	\\
3889 \hi+\hei	&	41.1	$\pm$	1.9	&	109.4	$\pm$	4.1	&	52.8	$\pm$	2.5	\\
3968 \neiii	&	19.1	$\pm$	1.6	&	59.9	$\pm$	3.6	&	23.8	$\pm$	2.2	\\
3970 \hepsi	&	28.7	$\pm$	1.7	&	83.4	$\pm$	3.6	&	41.6	$\pm$	2.3	\\
4102 \hdelta	&	45.5	$\pm$	1.7	&	134.5	$\pm$	3.2	&	67.3	$\pm$	1.9	\\
4341 \hgamma	&	56.6	$\pm$	0.7	&	150.5	$\pm$	2.2	&	56.1	$\pm$	0.5	\\
4363 \oiii	&	11.1	$\pm$	0.5	&	22.1	$\pm$	0.9	&	6.4	$\pm$	0.5	\\
4471 \hei	&	4.7	$\pm$	0.3	&	13.3	$\pm$	0.6	&	5.0	$\pm$	0.3	\\
4658 \feiii	&	0.9	$\pm$	0.2	&	3.8	$\pm$	0.6	&	1.6	$\pm$	0.3	\\
4686 \heii	&	1.1	$\pm$	0.3	&	2.6	$\pm$	0.6	&	0.4	$\pm$	0.2	\\
4711 \ariv	&	0.4	$\pm$	0.3	&	1.9	$\pm$	0.6	&	0.9	$\pm$	0.3	\\
4713 \hei	&	...	&	1.2	$\pm$	0.6	&	0.2	$\pm$	0.2	\\
4740 \ariv	&	1.7	$\pm$	0.5	&	1.8	$\pm$	0.5	&	0.6	$\pm$	0.3	\\
4861 \hbeta	&	126.6	$\pm$	0.9	&	352.7	$\pm$	2.6	&	140.4	$\pm$	0.7	\\
4922 \hei	&	1.3	$\pm$	0.4	&	3.3	$\pm$	0.5	&	1.2	$\pm$	0.3	\\
4959 \oiii	&	189.5$\pm$2.7	&	564.4$\pm$	3.4	&	214.0	$\pm$	1.2	\\
4988 \feiii	&	2.2$\pm$	0.2     &	5.5$\pm$	0.4 &	2.5$\pm$	0.3	\\
5007 \oiii	&	567.9	$\pm$	7.5	&	1673.2	$\pm$	14.8	&	646.8	$\pm$	3.3	\\
5016 \hei	&	1.8	$\pm$	0.3	&	5.7	$\pm$	0.5	&	2.5	$\pm$	0.4	\\
6302 \oi	&	3.3	$\pm$	0.2	&	11.0	$\pm$	0.4	&	5.3	$\pm$	0.3	\\
6312 \siii	&	1.5	$\pm$	0.3	&	4.4	$\pm$	0.4	&	1.9	$\pm$	0.2	\\
6364 \oi	&	1.6	$\pm$	0.2	&	4.3	$\pm$	0.4	&	2.2	$\pm$	0.3	\\
6547 \nii	&	2.6	$\pm$	0.9	&	7.7	$\pm$	3.9	&	4.2	$\pm$	1.5	\\
6563 \halpha	&	308.7	$\pm$	0.8	&	926.5	$\pm$	4.1	&	425.5	$\pm$	1.4	\\
6583 \nii	&	5.8	$\pm$	0.8	&	23.2	$\pm$	4.0	&	16.4	$\pm$	1.4	\\
6678 \hei	&	2.5	$\pm$	0.2	&	7.6	$\pm$	0.4	&	3.7	$\pm$	0.3	\\
6716 \sii	&	14.7	$\pm$	0.3	&	53.0	$\pm$	0.6	&	29.7	$\pm$	0.4	\\
6730 \sii	&	11.4	$\pm$	0.3	&	40.4	$\pm$	0.6	&	23.0	$\pm$	0.3	\\
7065 \hei	&	2.3	$\pm$	0.8	&	8.1	$\pm$	1.9	&	3.8	$\pm$	0.9	\\

\hline
$A_v$         & 0.13$\pm$0.11 & 0.33$\pm$0.18 & 0.1$\pm$0.18  \\ 
$Q\times100$           & 0.00 & 0.00 & 0.13  \\ 
\lum(\ulum)  & 40.43$\pm$0.11 & 40.97$\pm$0.13    & 40.47$\pm$0.11  \\

\eqwhbe      & 184$\pm$4 & 176$\pm$8 & 129$\pm$6  \\ 
\eqwhal      & 429$\pm$12 & 312$\pm$10 & 302$\pm$9  \\ 
$n_e$(\sii) (cm$^{-3}$)        & 202$\pm$86 & 149$\pm$ 72 & 152$\pm$	61  \\ 
$T_e$(\oiii)$\times10^{4}$ (K) & 1.62$\pm$	0.07 & 1.34$\pm$	0.05 & 1.15$\pm$	0.04  \\ 
\ohmas    & 7.39	$\pm$0.08 & 7.67	$\pm$0.07 & 8.01$\pm$	0.07  \\ 
\ohmasmas & 7.59	$\pm$0.06 & 7.55$\pm$	0.06 & 7.99$\pm$	0.05  \\ 
\oh       & 7.72	$\pm$0.06 & 7.91	$\pm$0.05 & 8.30$\pm$	0.06  \\ 
$\log$(N/O) & $-1.51\pm$ 0.04 & $-1.56\pm$0.04 & $-1.47\pm$0.05  \\ 
\he       & 10.80$\pm$ 0.03  & 10.81$\pm$0.03 & 10.83$\pm$0.03  \\ 
\hline
\multicolumn{4}{c}{Velocity dispersion measurements (\kms)}   \\
\hline
$\sigma$(\hbeta)            & 32.0$\pm$1.6	& 37.0$\pm$1.5 & 38.8$\pm$1.7\\
$\sigma$(\oiii$\lambda4959$)	 & 29.4$\pm$1.5	& 34.9$\pm$1.4 & 36.9$\pm$1.5\\
$\sigma$(\oiii$\lambda5007$) & 29.7$\pm$1.3	& 35.0$\pm$1.3 & 37.0$\pm$1.4\\
$\sigma$(\halpha)	         & 32.1$\pm$1.2	& 37.2$\pm$1.2 & 39.1$\pm$1.3\\
\hline 
\end{tabular} 
\end{table}


\end{document}